\documentclass[aps,preprintnumbers,superscriptaddress,runinaddress,amsmath,amssymb,showpacs,floatfix,twocolumn]{revtex4}

\usepackage{graphicx}
\usepackage{dcolumn}
\usepackage{bm}
\usepackage{amsmath}
\usepackage{hyperref} 
\usepackage{xcolor}

\begin{document}

\preprint{ADP-10-8/T704}

\title{Positive-parity Excited-states of the Nucleon in Quenched Lattice QCD}

\author{M. S. Mahbub}
\affiliation{Special Research Centre for the Subatomic Structure of Matter, Adelaide, South Australia 5005, Australia, \\ and Department of Physics, University of Adelaide, South Australia 5005, Australia.}\affiliation{Department of Physics, Rajshahi University, Rajshahi 6205, Bangladesh.}
\author{Alan $\acute{\rm{O}}$ Cais}
\affiliation{Special Research Centre for the Subatomic Structure of Matter, Adelaide, South Australia 5005, Australia, \\ and Department of Physics, University of Adelaide, South Australia 5005, Australia.}
\affiliation{Cyprus Institute, Guy Ourisson Builiding, Athalassa Campus, PO Box 27456, 1645 Nicosia, Cyprus.}
\author{Waseem Kamleh}
\author{Derek B. Leinweber}
\author{Anthony G. Williams}
\affiliation{Special Research Centre for the Subatomic Structure of Matter, Adelaide, South Australia 5005, Australia, \\ and Department of Physics, University of Adelaide, South Australia 5005, Australia.}

\collaboration{CSSM Lattice Collaboration}

\begin{abstract}

Positive-parity spin-$\frac{1}{2}$
excitations of the nucleon are explored in lattice QCD. The
variational method is used in this investigation and several
correlation matrices are employed. As our focus is on the utility and
methodology of the variational approach, we work in the quenched
approximation to QCD. Various sweeps of Gaussian
fermion-field smearing is applied at the source and at the sink of
$\chi_{1}\bar\chi_{1}$ and $\chi_{1}\chi_{2}$ correlation functions to
obtain a large basis of operators. Using several different approaches
for  constructing basis interpolators, we demonstrate how improving
the basis  can split what otherwise might be interpreted as a single
state  into multiple eigenstates. Consistency of the extracted excited
energy states are explored over various dimensions of the correlation
matrices. The use of large correlation matrices is
emphasized for the reliable extraction of the excited eigenstates of QCD.

\end{abstract}

\pacs{11.15.Ha,12.38.Gc,12.38.-t}

\maketitle

\section{ \label{sec1:intro}Introduction}

One of the long-standing puzzles in hadron spectroscopy has been the
low mass of the first positive parity, $J^{P}={\frac{1}{2}}^{+}$,
excitation of the nucleon, known as the Roper resonance $N^{*}$(1440
MeV). In constituent or valence quark models with harmonic oscillator
potentials, the lowest-lying odd parity state naturally occurs below
the $N={\frac{1}{2}}^{+}$ state (with principal quantum number $N=2$)
~\cite{Isgur:1977ef,Isgur:1978wd} whereas, in Nature the Roper
resonance is almost 100 MeV below the $N={\frac{1}{2}}^{-}$(1535 MeV)
state. Similar difficulties in the level orderings appear for the
$J^{P}={\frac{3}{2}}^{+} \Delta^{\ast}(1600)$ and ${\frac{1}{2}}^{+}
\Sigma^{\ast} (1690)$ resonances, which have led to the speculation
that the Roper resonance may be more appropriately viewed as a hybrid
baryon state with explicitly excited glue field configurations
~\cite{Li:1991yba,Carlson:1991tg} or as a breathing mode of the
ground state ~\cite{Guichon:1985ny} or states which can be described
in terms of meson-baryon dynamics alone ~\cite{Krehl:1999km}. 

Lattice QCD is very successful in computing many properties of hadrons
from first principles. In particular, in hadron spectroscopy, the
ground states of the hadron spectrum are now well
understood~\cite{Durr:2008zz}. However, the
excited states still prove a significant challenge as they belong to
the  sub-leading exponential of the two-point
correlation function. Extracting excited states masses from
these sub-leading exponents is difficult as the correlation
functions decay quickly and the signal to noise ratio 
deteriorates more rapidly. In baryon
spectroscopy, there are many experimentally observed baryon resonances
whose physical properties are poorly understood. Lattice QCD can
provide theoretical input to solidify their identification. The
first detailed analysis of the positive parity excitation of the nucleon
was performed in Ref.~\cite{Leinweber:1994nm} using Wilson fermions
and an operator product expansion spectral ansatz. Since then several
attempts have been made to address these issues in the lattice
framework~\cite{Lee:1998cx,Gockeler:2001db,Sasaki:2001nf,Melnitchouk:2002eg,Edwards:2003cd,Lee:2002gn,Mathur:2003zf,Sasaki:2003xc,Basak:2007kj,Bulava:2009jb,Bulava:2010yg,Engel:2010my},
but in many cases no potential identification of the Roper state has been made. Recently, however, in the analysis of
Refs.~\cite{Lee:2002gn,Mathur:2003zf,Sasaki:2005ap}
 a low-lying Roper state has been identified using Bayesian techniques.

Another state-of-the-art approach in hadron spectroscopy is the `variational
method' ~\cite{Michael:1985ne,Luscher:1990ck}, which is
based on a correlation matrix analysis. The identification of the
Roper state with this method have been mixed. However,
recently, in Ref.~\cite{Mahbub:2009aa} a low-lying Roper
state has been identified with this approach employing a diverse
range of smeared-smeared correlation functions. Our work there
motivates us to investigate the several positive parity excited states using
similar techniques but in a significantly more comprehensive manner.

In this paper, the variational analysis used in
Refs.~\cite{Mahbub:2009nr,Mahbub:2009aa} is explored more
extensively. In particular, we consider $6\times 6$ and  
$8\times 8$ correlation matrices not only built from the $\chi_{1}$
interpolating  field, but 
also incorporating another nucleon interpolator,
 $\chi_{2}$, to extend the set of basis operators. $6\times 6$
matrices are  built up using $\chi_{1}\bar\chi_{1}$ and
$\chi_{1}\chi_{2}$ correlators, while the $8\times8$ matrices use the
$\chi_{1}\chi_{2}$ correlation functions, as discussed in the text.

One of the goals of this paper is to investigate the high-lying
 positive parity spin-$\frac{1}{2}$ excited states of the nucleon, such as
$\rm{P}_{11}$(1710 MeV) and $\rm{P}_{11}$(2100 MeV), using larger
 correlation matrices. Incorporating the $\chi_{2}$
interpolator with various numbers of smearing sweeps, enables us to
explore more deeply  the overlapping of different interpolators with
the energy eigen-states. This will also prove the reliability
of the discovery of the Roper resonance~\cite{Mahbub:2009aa}. 
 We demonstrate how improving the basis interpolating fields can split what
otherwise might be interpreted as a single state into multiple
eigenstates.

This paper is arranged as follows:
Section~\ref{section:mass_of_hadrons} contains the general
description of the extraction of masses with the introduction of
different nucleon interpolating fields. The lattice details are given in
Section~\ref{section:simulation_details}, the results are
presented in Section~\ref{section:results}, and
conclusions are made in Section~\ref{section:conclusions}.     
%
%
\section{Mass of Hadrons}
\label{section:mass_of_hadrons}
The masses of hadrons are extracted from two-point correlation
functions using operators chosen to have overlap with desired
states. Let us consider a baryon state B of spin half, if we suppress
Dirac indices a two point function can be written as,  

\begin{align}
 {G_{ij}(t,\vec p)}&=\sum_{\vec x}e^{-i{\vec p}.{\vec x}}\langle{\Omega}\vert T \{ \chi_i (x)\bar\chi_j(0)\} \vert{\Omega}\rangle. 
\label{sec:mass:first_eqn}
\end{align}
The operator $\bar{\chi}_j(0)$ creates states from the vacuum at space-time
point $0$ and, following the evolution of the states in time $t$, the
states are destroyed by the operator $\chi_{i}(x)$ at point $\vec{x},t$. $T$
stands for the time ordered product of operators. 
A complete set of momentum eigenstates provides,

\begin{align}
\sum_{B,{\vec p}^{\, \prime},s}\vert{B,{\vec p}^{\,
    \prime},s}\rangle\langle{B,{\vec p}^{\, \prime},s}\vert &=I,
\label{sec:mass:completeness_eqn}
\end{align}
where $B$ can include multi-particle states. The substitution of Eq.~(\ref{sec:mass:completeness_eqn}) into Eq.~(\ref{sec:mass:first_eqn}) yields,

\begin{align}
{G_{ij}(t,\vec p)} &=\sum_{\vec x}\sum_{B,{\vec p}^{\,
    \prime},s}e^{-i{\vec p}.{\vec x}}\langle
{\Omega}\vert\chi_i(x)\vert{B,{\vec p}^{\,
    \prime},s}\rangle\langle{B,{\vec p}^{\, \prime},s}\vert\bar\chi_j(0)\vert {\Omega}\rangle.\label{sec:mass:MinkowskiCorrelationFunction_eqn1}
\end{align}
We can express the operator $\chi_{i}(x)$ as

\begin{align}
\chi_{i}(x) &= e^{iP.x}\chi_{i}(0)e^{-iP.x},
\end{align}
where, $P^{\mu}=P=(H,\vec{P})$ and $\vec{P}$ is the momentum operator whose eigenvalue is the total momentum of the system. Eq.~(\ref{sec:mass:MinkowskiCorrelationFunction_eqn1}) can now be written as, 

\begin{widetext}
\begin{align}
{G_{ij}(t,\vec p)} &=\sum_{\vec x}\sum_{B,{\vec p}^{\,
    \prime},s}e^{-i{\vec p}.{\vec x}}\langle
{\Omega}\vert{e^{iPx}\chi_{i}(0)e^{-iPx}}\vert{B,{\vec p}^{\,
    \prime},s}\rangle\langle{B,{\vec p}^{\, \prime},s}\vert\bar\chi_{j}(0)\vert {\Omega}\rangle \nonumber \\
&=\sum_{\vec x}\sum_{B,{\vec p}^{\, \prime},s}e^{-iE_{B}t}e^{-i{\vec
    x}.({\vec p}-{\vec p}^{\, \prime})}\langle
{\Omega}\vert\chi_{i}(0)\vert{B,{\vec p}^{\,
    \prime},s}\rangle\langle{B,{\vec p}^{\, \prime},s}\vert\bar\chi_{j}(0)\vert {\Omega}\rangle. 
\end{align}
\end{widetext}
As we move from Minkowski space to Euclidean space, the time $t\rightarrow {-it}$ and the above equation then can be written as,

\begin{align}
{G_{ij}(t,\vec p)} &=\sum_{B,{\vec p}^{\,
    \prime},s}e^{-E_{B}t}\delta_{\vec p,{\vec p}^{\, \prime}}\langle
{\Omega}\vert\chi_{i}(0)\vert{B,{\vec p}^{\,
    \prime},s}\rangle\langle{B,{\vec p}^{\, \prime},s}\vert\bar\chi_{j}(0)\vert {\Omega}\rangle\nonumber \\
&=\sum_{B}\sum_{s}e^{-E_{B}t}\langle {\Omega}\vert\chi_{i}(0)\vert{B,{\vec p},s}\rangle\langle{B,{\vec p},s}\vert\bar\chi_{j}(0)\vert {\Omega}\rangle.
\label{sec:mass:EucleadianCorrelationFunction_eqn1}
\end{align}
The overlap of the interpolating fields $\chi(0)$ and ${\bar\chi}(0)$ with positive and negative parity baryon states $\vert {B^{\pm}}\rangle$ can be parametrized by a complex quantity called the coupling strength, $\lambda_{B^{\pm}}$, which can be defined for positive parity states by

\begin{align}
\langle{\Omega}\vert\chi(0)\vert {B^{+}},\vec p,s\rangle &=\lambda_{B^{+}}\sqrt {\frac{M_{B^{+}}}{E_{B^{+}}}}u_{B^{+}}({\vec p},s),
\end{align}

\begin{align}
\langle B^{+},\vec p,s\vert\bar{\chi}(0)\vert {\Omega}\rangle &=\bar\lambda_{B^{+}}\sqrt {\frac{M_{B^{+}}}{E_{B^{+}}}}{{\bar u}_{B^{+}}}({\vec p},s).
\end{align}
For the negative parity states one requires

\begin{align}
\langle{\Omega}\vert\chi(0)\vert {B^{-}},\vec p,s\rangle &=\lambda_{B^{-}}\sqrt {\frac{M_{B^{-}}}{E_{B^{-}}}}\gamma_5{u_{B^{-}}({\vec p},s)},
\end{align}

\begin{align}
\langle B^{-},\vec p,s\vert\bar{\chi}(0)\vert {\Omega}\rangle &=-\bar\lambda_{B^{-}}\sqrt {\frac{M_{B^{-}}}{E_{B^{-}}}}{{\bar u}_{B^{-}}}({\vec p},s)\gamma_5.
\end{align}
Here, $\lambda_{B^{\pm}} $ and ${\bar\lambda}_{B^{\pm}}$ are the couplings of the interpolating functions at the sink and the source respectively and $M_{B^{\pm}}$ is the mass of the state $B^{\pm}$. ${E_{B^{\pm}}}$ is the energy of the state $B^{\pm}$, where ${E_{B^{\pm}}} = \sqrt{M^{2}_{B^{\pm}}+{\vec p}^2}$, and $u_{B^{\pm}}(\vec p,s)$ and ${{\bar u}_{B^{\pm}}}(\vec p,s)$ are the Dirac spinors, 

\begin{align}
{\bar u}^{\alpha}_{B^{\pm}}(\vec p,s) {u^{\beta}_{B^{\pm}}}(\vec p,s) &= \delta{^{\alpha \beta}}.
\end{align}
Thus, Eq.~(\ref{sec:mass:EucleadianCorrelationFunction_eqn1}) contains a projection operator $\Gamma_{\pm}=\sum_{s} {u^{\beta}_{B^{\pm}}}(\vec p,s){\bar u}^{\alpha}_{B^{\pm}}(\vec p,s)$, through which the contributions to the even and odd parity states from the correlation function can be obtained. For positive parity, this can be expressed as, 

\begin{align} 
\sum_{s} {u^{\beta}_{B^{+}}}(\vec p,s){\bar u}^{\alpha}_{B^{+}}(\vec p,s) &=\frac{\gamma .p + M_{B^{+}}}{2{E_{B^{+}}}},
\end{align}

and for the negative parity,

\begin{align} 
\gamma_{5}\left(\sum_{s} {u^{\beta}_{B^{-}}}(\vec p,s){\bar u}^{\alpha}_{B^{-}}(\vec p,s)\right)\gamma_{5} &=\frac{-\gamma .p + M_{B^{-}}}{2{E_{B^{-}}}}.
\end{align}
By substituting the above Eqs. for the positive and negative parity states in Eq.~(\ref{sec:mass:EucleadianCorrelationFunction_eqn1}) we obtain,
\begin{align}
{\cal{G}}_{ij}(t,\vec p) &=\sum_{B^{+}}\lambda_{B^{+}}\bar\lambda_{B^{+}}e^{{-E_{B^{+}}}t} {\frac{\gamma .p_{B^{+}} + M_{B^{+}}}{2E_{B^{+}}}} \nonumber \\
              & +\sum_{B^{-}}\lambda_{B^{-}}\bar\lambda_{B^{-}}e^{{-E_{B^{-}}}t} {\frac{-\gamma .p_{B^{-}} + M_{B^{-}}}{2E_{B^{-}}}}.\label{sec:mass:FinalCorrelationFunction_eqn}
\end{align}
At momentum $\vec p=\vec 0$, $E_{B^{\pm}}=M_{B^{\pm}}$, a parity projection operator $\Gamma_{\pm}$ can be introduced,

\begin{align}
\Gamma_{\pm} &= \frac{1}{2}(1\pm \gamma_0).
\end{align}
 We can isolate the masses of the even and odd parity states by taking the trace of $\cal{G}$ with the operators $\Gamma_{+}$ and $\Gamma_{-}$. The positive parity state propagates through the (1,1) and (2,2) elements of the Dirac matrix, whereas, negative parity state propagates through the (3,3) and (4,4) elements.

The correlation function for positive and negative parity states can then be written as,

\begin{align}
G_{ij}^{\pm}(t,\vec 0) &= {\rm{Tr}}_{\rm sp}[\Gamma_{\pm}{\cal{G}}_{ij}(t,\vec 0)]\nonumber \\
&= \sum_{B^{\pm}}\lambda_{i}^{\pm}\bar\lambda_{j}^{\pm}e^{{-M_{B^{\pm}}}t}.
\end{align}
The correlation function contains a superposition of states. The mass of the lowest state, $M_{0^{\pm}}$ can be extracted at large $t$ where the contributions from all other states are suppressed,

\begin{align}
G_{ij}^{\pm}(t,\vec 0) & \stackrel{t \rightarrow \infty}{=} \lambda_{i0}^{\pm}\bar\lambda_{j0}^{\pm}e^{{-M_{0^{\pm}}}t}.
\end{align}

\subsection{Source Smearing}
The spatial source smearing~\cite{Gusken:1989qx} technique is applied to
increase the overlap of the interpolators with the lower lying
states. We employ a fixed boundary condition in the time direction 
for the fermions by setting $U_t(\vec x,N_t)=0\,\forall\,{\vec x}$ in
the hopping terms of the fermion action with periodic boundary
conditions imposed in the spatial directions. Gauge invariant Gaussian
smearing ~\cite{Gusken:1989qx} in the spatial dimensions is applied
through an iterative process. The smearing procedure is: 

\begin{align}
\psi_{i}(x,t) &=\sum_{x'}F(x,x')\psi_{i-1}(x',t),
\end{align}
where,
\begin{align}
F(x,x') &= {(1-\alpha)}\delta_{x,x'}+\frac{\alpha}{6}\sum_{\mu=1}^{3}[U_{\mu}(x)\delta_{x',x+\hat\mu} \nonumber \\
        & +U_{\mu}^{\dagger}(x-\hat\mu)\delta_{x',x-\hat\mu}],
\end{align}
where the parameter $\alpha=0.7$ is used in our calculation. After repeating the procedures $N_{\rm sm}$ times on a point source the resulting smeared fermion field is,
\begin{align}
\psi_{N_{\rm sm}}(x,t) &=\sum_{x'}F^{N_{sm}}(x,x')\psi_{0}(x',t).
\end{align}


\subsection{Variational Method}

The extraction of the ground state mass can be done
straightforwardly. However access to the excited state masses requires
additional effort. Here we consider the variational method
~\cite{Michael:1985ne,Luscher:1990ck}. The variational
method requires the cross correlation of operators so that the
operator space can be diagonalised and the excited state masses
extracted from the exponential nature of the diagonalised basis. To
access $N$ states of the spectrum, one requires a minimum of $N$
interpolators. With the assumption that only $N$ states contribute
significantly to $G_{ij}$ at time $t$, the parity projected two point
correlation function matrix for $\vec{p} =0$ can be written as, 

\begin{align}
G^{\pm}_{ij}(t) &= (\sum_{\vec x}{\rm Tr}_{\rm sp}\{ \Gamma_{\pm}\langle\Omega\vert\chi_{i}(x)\bar\chi_{j}(0)\vert\Omega\rangle\}) \\
          &=\sum_{\alpha=0}^{N-1}\lambda_{i}^{\alpha}\bar\lambda_{j}^{\alpha}e^{-m_{\alpha}t},
\end{align}
where Dirac indices are implicit. Here, $\lambda_{i}^{\alpha}$ and $\bar\lambda_{j}^{\alpha}$ are the
couplings of interpolators $\chi_{i}$ and  $\bar\chi_{j}$ at the sink
and source respectively to eigenstates $\alpha=0, \cdots
,(N-1)$. $m_{\alpha}$ is the mass of the state $\alpha$. The $N$
interpolators have the same quantum numbers and provide an
$N$-dimensional basis upon which to describe the states. Using this
basis we aim to construct $N$ independent interpolating source and
sink fields which isolate $N$ baryon states $\vert B_{\alpha}\rangle,$
{\it i.e.}

\begin{align}
{\bar\phi}^{\alpha} &=\sum_{i=1}^{N}u_{i}^{\alpha}{\bar\chi}_{i},
\end{align}   
\vspace{-0.60cm}
\begin{align}
{\phi}^{\alpha} &=\sum_{i=1}^{N}v_{i}^{\alpha}{\chi}_{i},
\end{align} 
such that,

\begin{align}
\langle{B_{\beta},p,s}\vert {\bar\phi}^{\alpha}\vert\Omega\rangle &= \delta_{\alpha\beta}{\bar{z}}^{\alpha}\bar{u}(\alpha,p,s),
\end{align}
\vspace{-0.60cm}
\begin{align}
\langle\Omega\vert{\phi}^{\alpha}\vert B_{\beta},p,s\rangle &= \delta_{\alpha\beta}{z}^{\alpha}u(\alpha,p,s),
\end{align}
where $z^{\alpha}$ and ${\bar{z}}^{\alpha}$ are the coupling strengths
of $\phi^{\alpha}$ and ${\bar\phi}^{\alpha}$ to the state $\vert
B_{\alpha}\rangle$. Consider a real eigenvector $u_{j}^{\alpha}$ which
operates on the correlation matrix $G_{ij}(t)$ from the right, one can
obtain ~\cite{Melnitchouk:2002eg}, 

\begin{align}
G_{ij}(t)u_{j}^{\alpha} &=(\sum_{\vec x}{\rm Tr}_{\rm sp}\{ \Gamma_{\pm}\langle\Omega\vert\chi_{i}\bar\chi_{j}\vert\Omega\rangle\})u_{j}^{\alpha} \nonumber \\
& = \lambda_{i}^{\alpha}\bar{z}^{\alpha}e^{-m_{\alpha}t}.
\end{align}
For notational convenience, in the remainder of the discussion the
repeated indices $i,j,k$ are to be understood as being summed over,
whereas, $\alpha$, which stands for a particular state, is not. Since
the only $t$ dependence comes from the exponential term, we can write
a recurrence relation at time $(t+\triangle t)$ as, 
\begin{align}
G_{ij}(t+\triangle t)u_{j}^{\alpha} & = e^{-m_{\alpha}\triangle t} G_{ij}(t)u_{j}^{\alpha},
\end{align}  
for sufficiently large $t$ and $t+\triangle t$~\cite{Blossier:2009kd,Mahbub:2009nr}.

Multiplying the above equation by $[G_{ij}(t)]^{-1}$ from the left we get,

\begin{align}
[(G(t))^{-1}G(t+\triangle t)]u^{\alpha} & = e^{-m_{\alpha}\triangle t}u^{\alpha} \nonumber \\
& = c^{\alpha}u^{\alpha}.
\label{eqn:right_eigenvalue_equation}
\end{align} 
This is an eigenvalue equation for eigenvector $u^{\alpha}$ with eigenvalue $c^{\alpha}=e^{-m_{\alpha}\triangle t}$. We can also solve the left eigenvalue equation to recover the $v^{\alpha}$ eigenvector,

\begin{align}
v_{i}^{\alpha}G_{ij}(t+\triangle t) & = e^{-m_{\alpha}\triangle t}v_{i}^{\alpha}G_{ij}(t).
\end{align} 
Similarly,

\begin{align}
v^{\alpha}[G(t+\triangle t)(G(t))^{-1}] & = e^{-m_{\alpha}\triangle t}v^{\alpha}.
\label{eqn:left_eigenvalue_equation}
\end{align} 
The vectors $u_{j}^{\alpha}$ and $v_{i}^{\alpha}$  diagonalize the correlation matrix at time $t$ and $t+\triangle t$ making the projected correlation matrix,

\begin{align}
v_{i}^{\alpha}G_{ij}(t)u_{j}^{\beta} = \delta^{\alpha\beta}z^{\alpha}{\bar{z}}^{\beta}e^{-m_{\alpha}t}.
 \label{eqn:projected_cf} 
\end{align} 
The parity projected, eigenstate projected correlator, $v_{i}^{\alpha}G^{\pm}_{ij}(t)u_{j}^{\alpha} \equiv G_{\pm}^{\alpha}$ is then used to obtain masses of different states.
We construct the effective mass 
\begin{align}
M_{\rm eff}^{\alpha}(t) &= {\rm ln}\left(\frac{{G_{\pm}^{\alpha}}(t,\vec 0)}{G_{\pm}^{\alpha}(t+1,\vec 0)}\right)\nonumber \\
 & = M_{\pm}^{\alpha}.
 \label{eqn:efective_mass}
\end{align}
and apply standardised analysis techniques as described in Ref.~\cite{Mahbub:2009nr}.
%
%
%
\section{Simulation details}
\label{section:simulation_details}
We use an ensemble of 200 quenched configurations with a lattice
volume of $16^{3}\times 32.$ Gauge field configurations are generated
by using the doubly blocked Wilson action in two coupling space
(DBW2)~\cite{Takaishi:1996xj,deForcrand:1999bi}. An
${\cal{O}}(a)$-improved fat link irrelevant clover (FLIC) fermion
action ~\cite{Zanotti:2001yb} is
used to generate quark propagators. This action has excellent scaling
properties and provides near continuum results at finite lattice
spacing ~\cite{Zanotti:2004dr}. The lattice spacing is $a=0.127$ fm,
as determined by the static quark potential, with the scale set with
the Sommer scale, $r_{o}=0.49$ fm ~\cite{Sommer:1993ce}. In the
irrelevant operators of the fermion action we apply four sweeps of
stout-link smearing to the gauge links to reduce the coupling with the
high frequency modes of the theory ~\cite{Morningstar:2003gk}. We use
the same method as in Ref.~\cite{Lasscock:2005kx} to determine fixed
boundary effects, and the effects are significant only after time
slice 25 in the present analysis. 

Eight different levels 
of gauge invariant Gaussian smearing~\cite{Gusken:1989qx} (1,
3, 7, 12, 16, 26, 35, 48 sweeps corresponding to rms 
  radii, in lattice units, of 0.6897, 1.0459, 1.5831, 2.0639, 2.3792,
  3.0284, 3.5237, 4.1868) are applied at the source (at $t=4$)
and at the sink. This is to ensure a variety of overlaps of the
interpolators with the lower-lying states. The analysis is performed
on nine different quark masses providing pion masses of
${m_{\pi}}$=\{0.797,0.729,0.641,0.541,0.430,0.380,0.327,0.295,0.249
\} GeV. The error analysis is performed using the jackknife method,
where the ${\chi^{2}}/{\rm{dof}}$ is obtained via a covariance matrix
analysis method.

The nucleon interpolators we consider in this analysis are
\begin{align}
\chi_1(x) &= \epsilon^{abc}(u^{Ta}(x)C{\gamma_5}d^b(x))u^{c}(x)\, , \\
\chi_2(x) &= \epsilon^{abc}(u^{Ta}(x)Cd^b(x)){\gamma_5}u^{c}(x)\, .
\end{align}
 We use the
Dirac representation of the gamma matrices in our analysis.

   Each of the $6\times 6$ correlation
   matrices of $\chi_{1}\bar\chi_{1}$ correlators corresponds to a
   particular selection of 6 levels of smearing from the 8 that we have
   available. We considered six different combinations of these to
   give six different correlation matrices.  Our
 selections are shown in Table~\ref{table:6x6_bases_x1x1}.  We note
 that 48  sweeps tends to be
 noisy and therefore eliminate it from most of our considerations.
  
For the $6\times 6$ matrices of $\chi_{1}\chi_{2}$ correlators, a
 subset of six bases is considered, as shown in
 Table~\ref{table:6x6_bases_x1x2} corresponding to a
   choice of 3 smearings to $\chi_{1}$ and the same 3 for $\chi_{2}$.
   For the $8\times 8$ matrices of $\chi_{1}\chi_{2}$ correlators, a subset of
 seven bases are considered and are given in
 Table~\ref{table:8x8_bases_x1x2} corresponding to a
   choice of 4 smearings to $\chi_{1}$ and the same 4 for $\chi_{2}$.
   The correlation matrices for
   $\chi_{1},\chi_{2}$ interpolators contain
 all combinations of correlation functions of $\chi_{1},\chi_{2}$,
 i.e.\ $\chi_{1}\bar\chi_{1}$, $\chi_{1}\bar\chi_{2}$,
 $\chi_{2}\bar\chi_{1}$ and $\chi_{2}\bar\chi_{2}$.

 It is noted that basis operators that are linearly dependent will cause the
  eigenvalue analysis to fail as there will be a singularity in the
  correlation matrix.  The fact that our analysis ($\chi_{1}\bar\chi_{1}$ and
  $\chi_{1}\chi_{2}$) succeeds
  indicates that our choices of operators access an equal number of dimensions
  in the Hilbert space.  Thus it is interesting to examine the stability of
  the masses to different choices of bases to ascertain whether one
  has  reliably isolated single eigenstates
  of QCD.  The relevant issues are: (i) whether or not the
  operators are sufficiently far from collinear that numerical errors
  do not prevent diagonalisation of the correlation matrix and, (ii)
  whether or not the states of interest have significant overlap with
  the subspace spanned by our chosen sets of
  operators. Since our correlation matrix
    diagonalisation succeeded, except at large Euclidean times where
    statistical errors dominate, we conclude that our operators
    are sufficiently far from collinear.

  \begin{table}[h]
    \begin{center}
       \caption{\label{table:6x6_bases_x1x1}The bases of $6\times 6$
         correlation matrices of $\chi_{1}\bar\chi_{1}$.} 
 \vspace{0.25cm}
     \begin{ruledtabular}
     \begin{tabular}{c|cccccccc} 
     Sweeps $\rightarrow$  & 1 & 3 & 7 & 12 & 16 & 26 & 35 & 48  \\
        \hline
   Basis No. $\downarrow$ & \multicolumn{8}{c}{Bases}  \\
        \hline    
    1 & 1 & 3 & 7 & 12 & 16 & 26 & - & - \\
    2 & 1 & 3 & 7 & 12 & 16 & - & 35 & - \\
    3 & 1 & 3 & 7 & - & 16 & 26 & 35 & -  \\
    4 & 1 & 3 & - & 12 & 16 & 26 & - & 48  \\ 
    5 & 1 & - & 7 & 12 & 16 & 26 & 35 & - \\  
    6 & - & 3 & 7 & 12 & 16 & 26 & 35 & - \\
 
 \end{tabular}
     \end{ruledtabular}
 \end{center}
 \end{table}

 \begin{table}[h]
    \begin{center}
 \caption{\label{table:6x6_bases_x1x2}The bases of $6\times 6$ correlation matrices of $\chi_{1}\chi_{2}$.}
\vspace{0.25cm}
     \begin{ruledtabular}
     \begin{tabular}{c|cccccccc} 
     Sweeps $\rightarrow$  & 1 & 3 & 7 & 12 & 16 & 26 & 35 & 48  \\
        \hline
   Basis No. $\downarrow$ & \multicolumn{8}{c}{Bases}  \\
        \hline    
     1 &  1 & - & - & - & 16 & - & - & 48   \\
     2 &  - & 3 & - & 12 & - & 26 & - & -   \\
     3 &  - & 3 & - & - & 16 & - & - & 48   \\
     4 &  - & - & 7 & - & 16 & - & 35 & -    \\ 
     5 &  - & - & - & 12 & 16 & 26 & - & -  \\  
     6 &  - & - & - & - & 16 & 26 & 35 & - \\
 
 \end{tabular}
     \end{ruledtabular}
 \end{center}
 \end{table}

  \begin{table}[h]
    \begin{center}
 \caption{\label{table:8x8_bases_x1x2}The bases of $8\times 8$ correlation matrices of $\chi_{1}\chi_{2}$.}
\vspace{0.25cm}
     \begin{ruledtabular}  
     \begin{tabular}{c|cccccccc} 
     Sweeps $\rightarrow$  & 1 & 3 & 7 & 12 & 16 & 26 & 35 & 48  \\
        \hline
   Basis No. $\downarrow$ & \multicolumn{8}{c}{Bases}  \\
        \hline    
    1 &  1 & - & 7 & -  & 16 & -  & 35 & -    \\
    2 &  - & - & 7 & 12 & 16 & 26 & -  & - \\
    3 &  - & 3 & - & 12 & -  & 26 & -  & 48   \\
    4 &  - & - & 7 & 12 & -  & 26 & 35 & -  \\ 
    5 &  - & - & 7 & -  & 16 & 26 & 35 & -  \\  
    6 &  - & - & 7 & -  & 16 & -  & 35 & 48  \\
    7 &  - & - & - & 12 & 16 & 26 & 35 & -  \\
 
 \end{tabular}
     \end{ruledtabular}  
 \end{center}
 \end{table}

\section{Results}
\label{section:results}
%

\subsection{Variational Analysis}

\begin{figure*}[!hpt]
  \begin{center}
   $\begin{array}{c@{\hspace{0.15cm}}c}  
 \includegraphics [height=0.35\textwidth,angle=90]{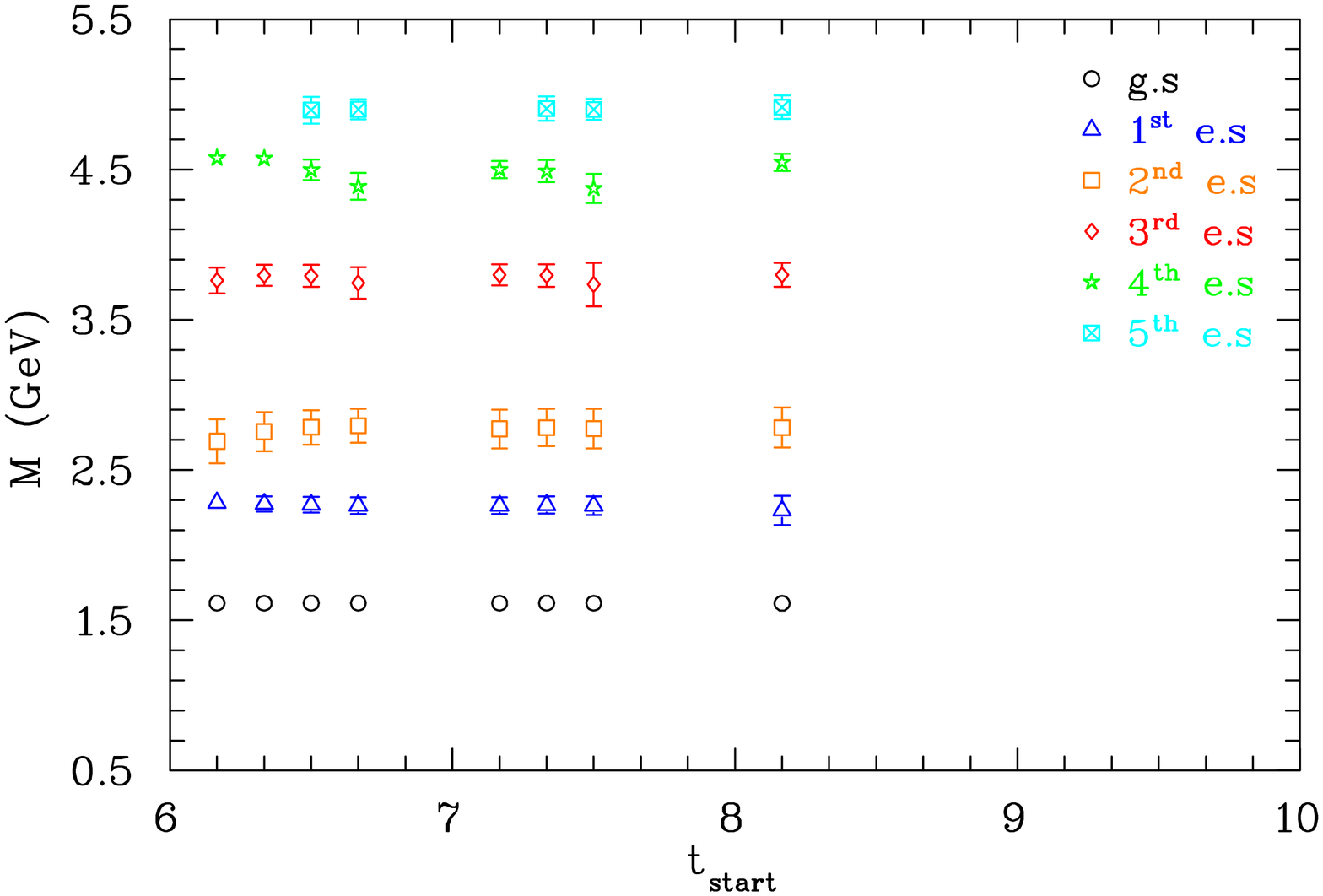} &
 \includegraphics [height=0.35\textwidth,angle=90]{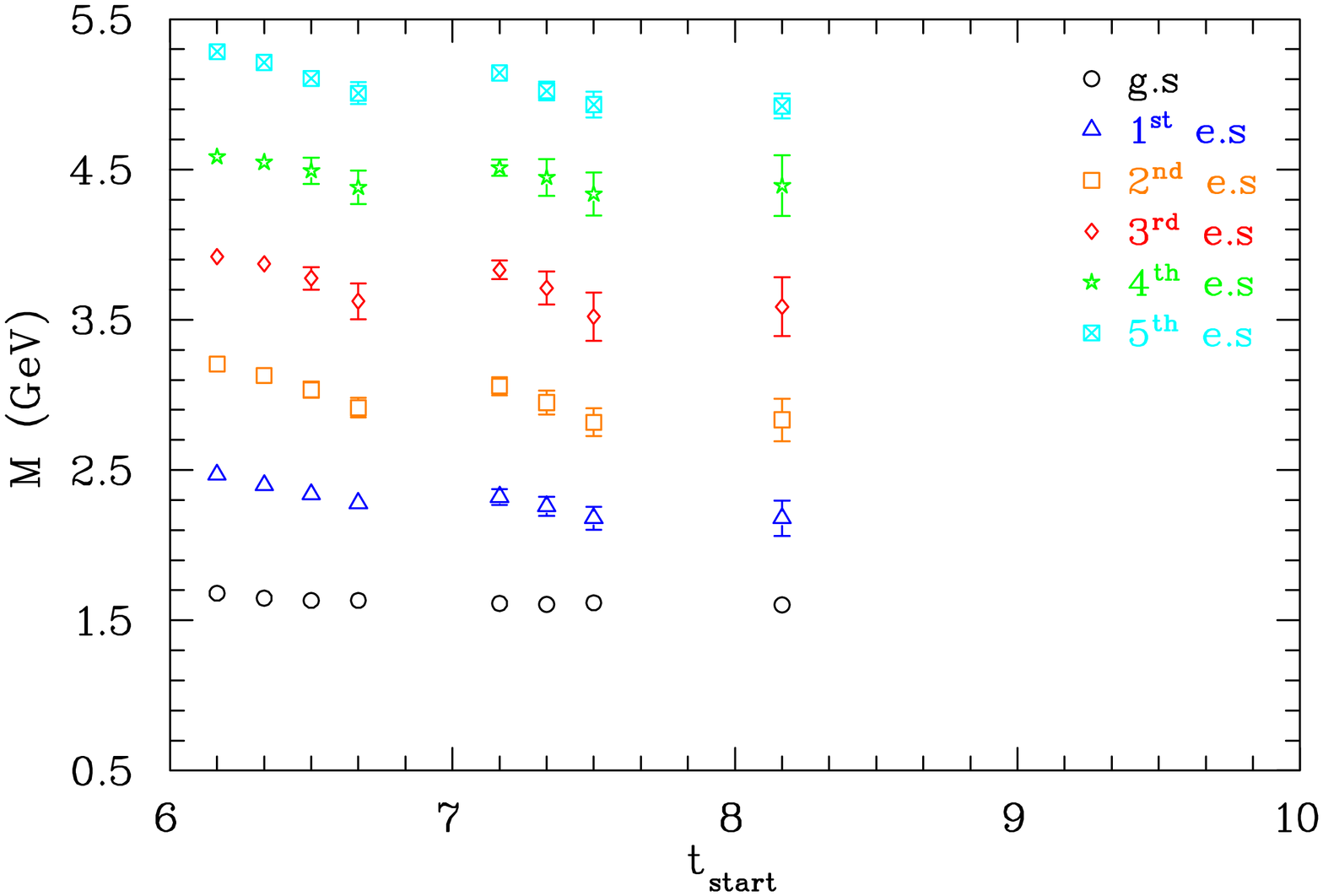}    
    \end{array}$

   $\begin{array}{c@{\hspace{0.15cm}}c}  
 \includegraphics [height=0.35\textwidth,angle=90]{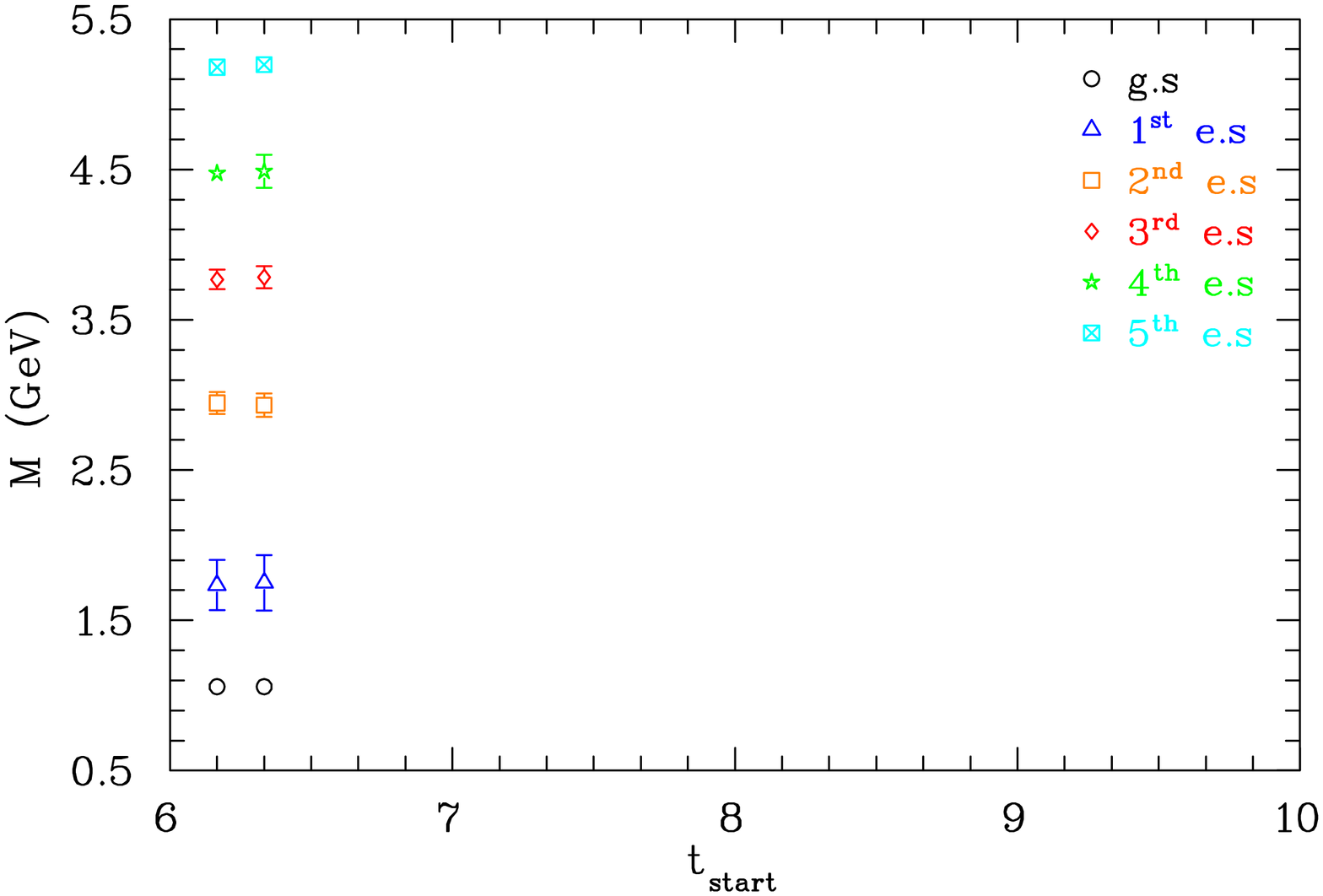} &
 \includegraphics [height=0.35\textwidth,angle=90]{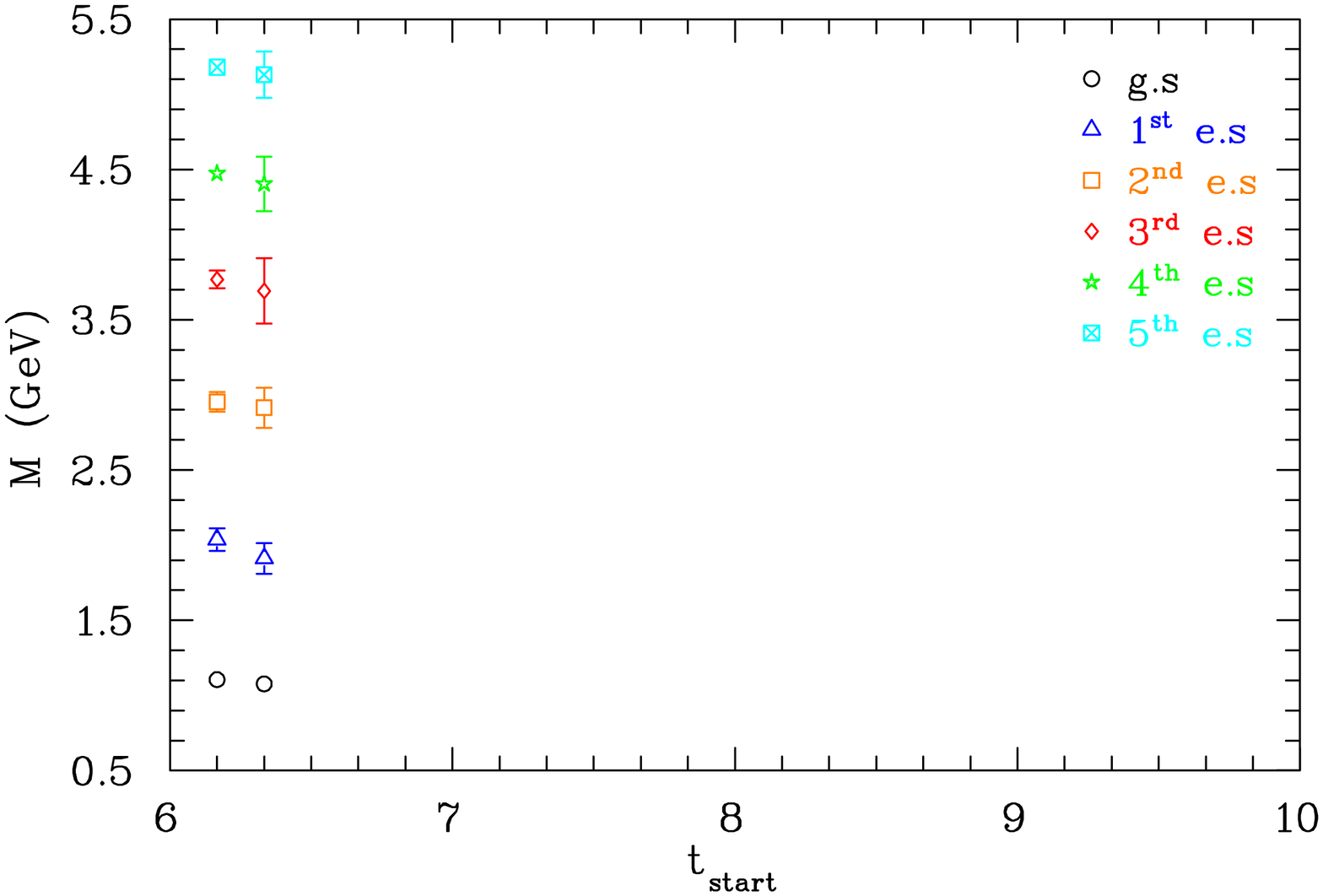}    
    \end{array}$

    \caption{(color online). Masses of the nucleon,
      $N^{{\frac{1}{2}}^{+}}$ states, from the projected correlation
      functions as shown in Eq.~\ref{eqn:projected_cf} (left) and  from
      the eigenvalues (right) for the $3^{\rm rd}$ combination
      (1,3,7,16,26,35 sweeps) of $6\times 6$ correlation matrices of
      $\chi_{1}\bar\chi_{1}$ correlation functions. The figure corresponds to
      pion masses of 797 MeV (top row) and 249 MeV (bottom row).
       Each pair of ground and excited state masses correspond to the
      diagonalization of the correlation matrix for each set of
      variational parameters $t_{\rm start}$ (shown in major tick marks) and
      $\triangle t$ (shown in minor tick marks). Here, $t_{\rm
        start}\equiv t$, as shown in
      Eqs.~(\ref{eqn:right_eigenvalue_equation}) and~(\ref{eqn:left_eigenvalue_equation}).}   
   \label{fig:mass_and_eig_for_x1x1_6x6}
  \end{center}
\end{figure*}

We begin by considering the $6\times 6$ correlation matrices of
$\chi_{1}\bar\chi_{1}$ 
correlators. In
Fig.~\ref{fig:mass_and_eig_for_x1x1_6x6}, masses
from the projected correlation functions and eigenvalues for the
$3^{\rm rd}$ basis of Table~\ref{table:6x6_bases_x1x1} are
presented. The basis elements for the $6\times 6$ matrices are highly linearly
dependent, and this means the analysis is more complicated than for
the $4\times 4$ matrices of Ref.~\cite{Mahbub:2009aa}. 
While the masses from the eigenvalues again display larger dependency
on the variational parameters, as observed in Refs.~\cite{Mahbub:2009nr,Mahbub:2009aa}, masses from the
projected correlation functions are very consistent on $t_{\rm start}$
($\equiv t$)
and $\triangle t$. Notably, the masses for the ground and first excited
states are as robust as in Ref.~\cite{Mahbub:2009aa}. For the fifth excited
state at the heaviest pion mass with the  
variational parameters ($t_{\rm start},\triangle t$) = (6,1),(6,2) and
(7,1), (top left graph of  Fig.~\ref{fig:mass_and_eig_for_x1x1_6x6}),
acceptable fits~\cite{Mahbub:2009nr} were unobtainable.  Nonetheless, the consistency
of the calculated masses over the significant sets of variational parameters 
 is self evident, as is how a mass can be exposed using one set of
 $t_{\rm start}$ and $\triangle t$.

From a series of $t_{\rm start}$
 and $\triangle t$, a single mass is 
selected for one set of $t_{\rm start}$ and $\triangle t$ by the
selection criteria discussed in Ref.~\cite{Mahbub:2009nr}, where we
prefer larger value of $t_{\rm start}+\triangle
t$~\cite{Blossier:2009kd}. In cases where a
larger $t_{\rm start}+\triangle t$ provides a poor signal-to-noise
ratio, for example ($t_{\rm start},\triangle t$)=(7,3) (top left
figure), we prefer a little lower $t_{\rm start}+\triangle t$ value,
for example ($t_{\rm start},\triangle t$)=(7,2), and we follow this
procedure for each quark mass.

\begin{figure*}[!hpt]
  \begin{center}
   $\begin{array}{c@{\hspace{0.15cm}}c}  
 \includegraphics [height=0.40\textwidth,angle=90]{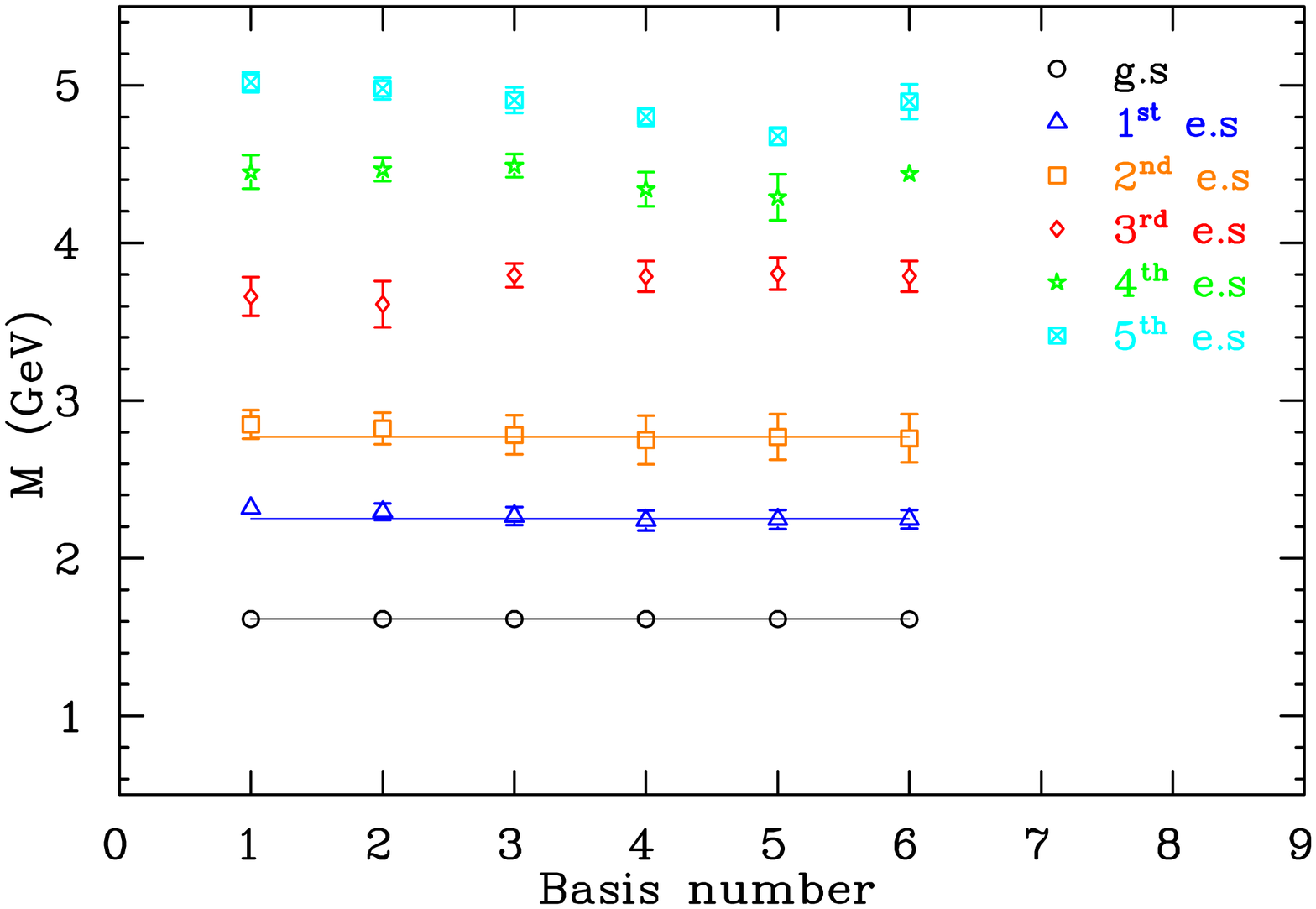} &
 \includegraphics [height=0.40\textwidth,angle=90]{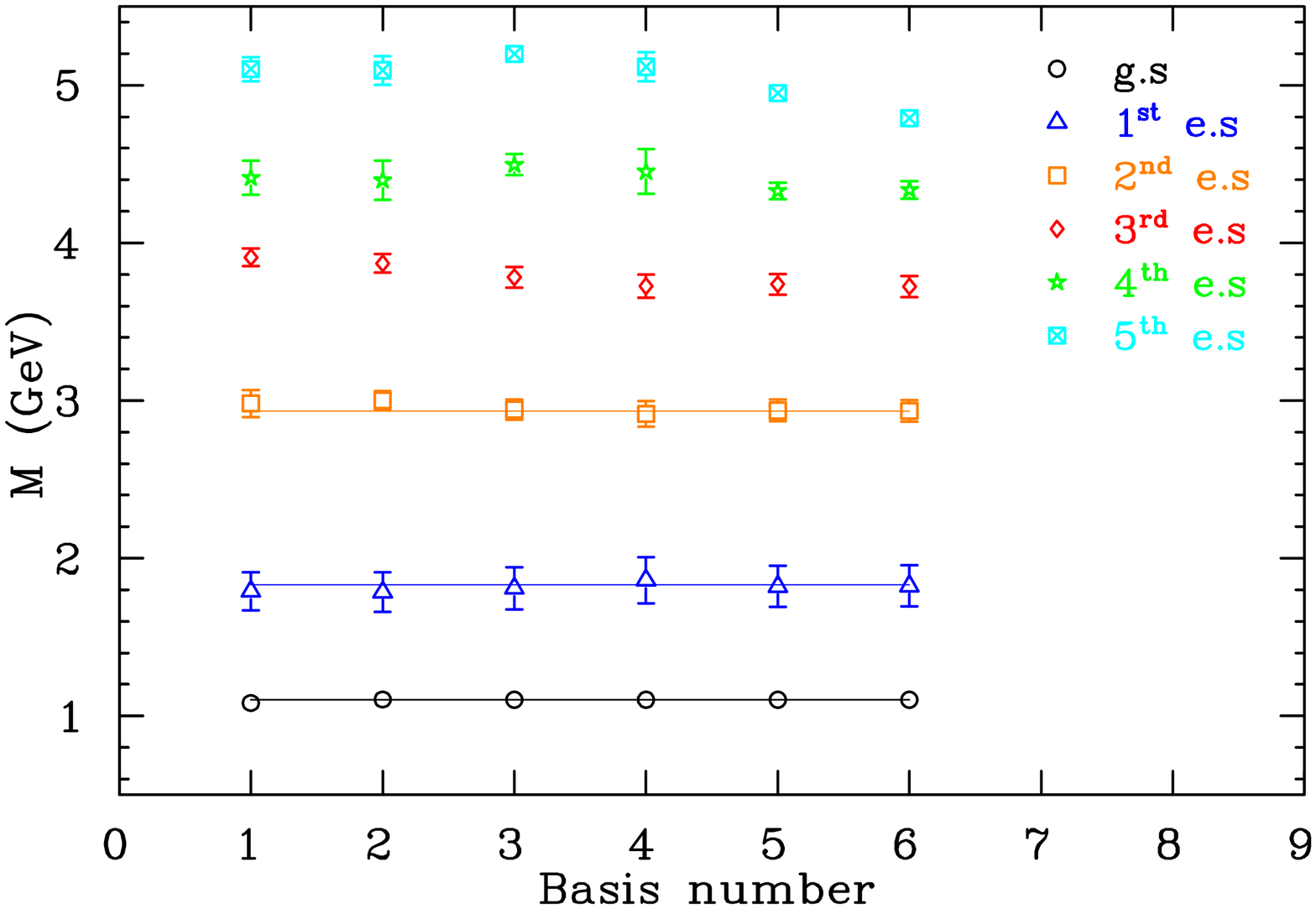}    
    \end{array}$
    \caption{(color online). Masses of the nucleon, $N^{{\frac
          {1}{2}}^{+}}$ states, from the projected correlation functions for
       the $6\times 6$ correlation matrices of $\chi_{1}\bar\chi_{1}$
       correlators, for pion masses of 797 MeV (left) and 295 MeV
       (right).  Numbers on
      the horizontal scale correspond to each basis of $6\times 6$
      matrices, for instance, 1 and 2 correspond to the bases
      of (1,3,7,12,16,26 sweeps) and (1,3,7,12,16,35
      sweeps) respectively, and so 
      on. Masses are extracted according to the selection criteria described in
      the text and in Ref.~\cite{Mahbub:2009nr}
      from all the  
      combinations of $6\times 6$ correlation matrices as shown in
      Table~\ref{table:6x6_bases_x1x1}. Straight lines are drawn to
      illustrate the  invariance of
the masses over the bases.}   
   \label{fig:mass_6x6_x1x1_all_combinations}
  \end{center}
\end{figure*}

\begin{figure*}[!hpt]
  \begin{center}

 \includegraphics [height=0.50\textwidth,angle=90]{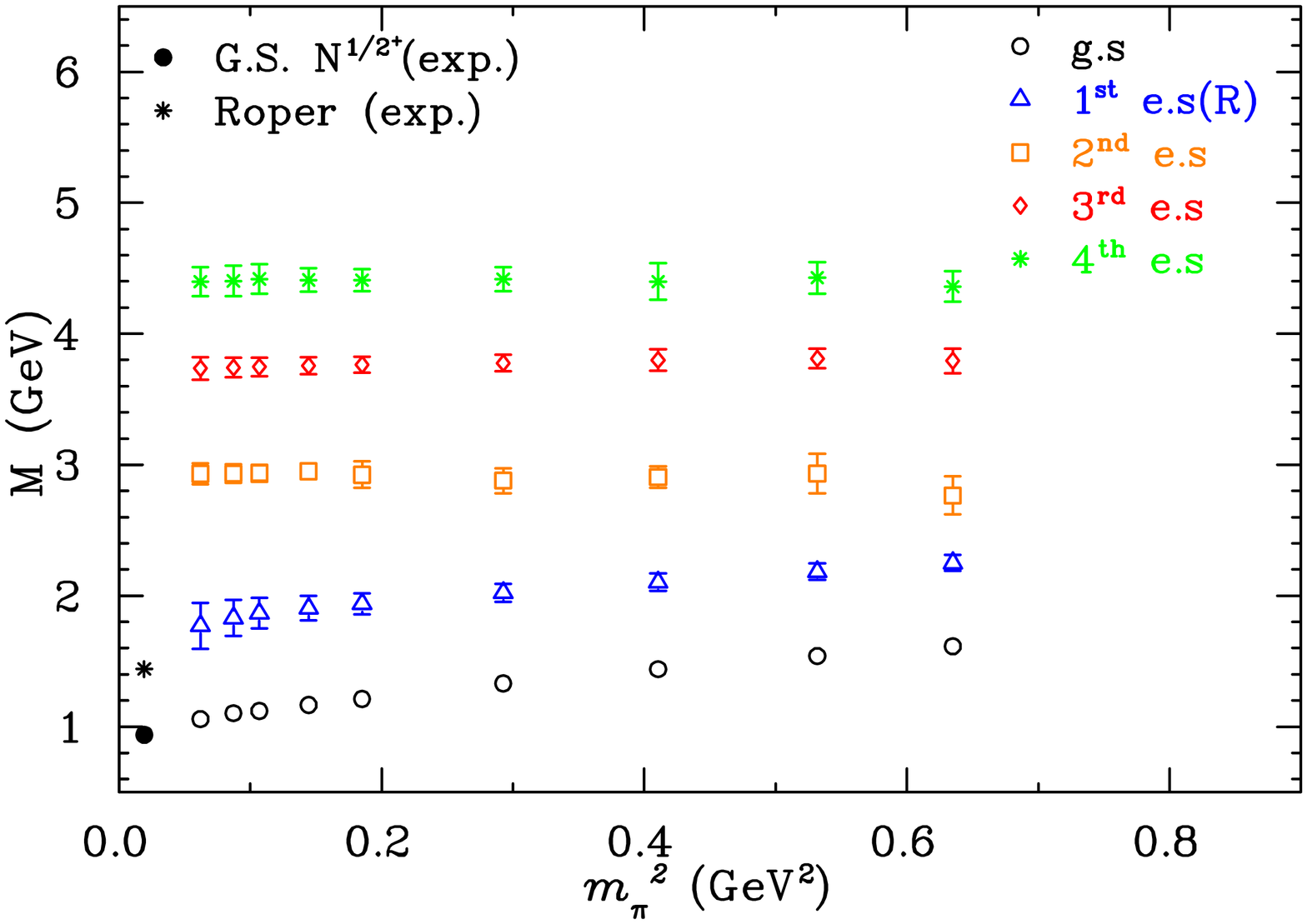}

    \caption{(color online). Masses of the nucleon,
      $N^{{\frac{1}{2}}^{+}}$ states, for the ground (g.s) and the
      excited (e.s) states from the
       $6\times 6$ correlation matrices of $\chi_{1}\bar\chi_{1}$
      correlators over four bases  (from the $3^{\rm rd}$ to the
      $6^{\rm th}$ of Fig.~\ref{fig:mass_6x6_x1x1_all_combinations}). The
      errors shown in the figure are a combination of average statistical errors
       over these four bases and systematic errors due to basis
      choices. Errors
      are combined in quadrature. The
      black filled symbols are the experimental values of the ground
      and the Roper states obtained from Ref.~\cite{Amsler:2008zzb}.
       Masses are given in Table~\ref{table:mpi.mavg.err-sum-over-qudrature.4comb.6states.x1x1}.}  
   \label{fig:m.avg_x1x1.6x6.sqrt.avgstaterr.basis_4combs.3rdTo6thCombs.5states.allQ}
  \end{center}
\end{figure*}

In Fig.~\ref{fig:mass_6x6_x1x1_all_combinations}, the masses for all the
$6\times 6$ correlation matrices of $\chi_{1}\bar\chi_{1}$ correlators are
presented. Straight lines are drawn to illustrate the invariance of
the masses over the bases. Since the $1^{\rm st}$ basis contains
all the  consecutive smearing sweep counts, i.e. 1,3,7,12,16,26 sweeps, it
may not span the space as well as other choices. This
basis  exposes  higher first and second excited states and this remains
 true for the $2^{\rm nd}$ basis where it contains
consecutive five lower smearing sweeps (left figure). These are therefore
 less reliable bases for extracting excited state masses. It is noted
 that  masses 
 from the third excited state onwards contain higher
 fluctuations than the lower states, independent of the choice of
 basis. A basis dependence is also evident for the third
 excited state (right figure).
 However, the bases from the $3^{\rm rd}$ to the $6^{\rm th}$ sets are
 more consistent, and in particular, the lower three
states are stable over these four bases. Therefore,
 the bases from $3^{\rm rd}$ to $6^{\rm th}$ are
used to  perform a systematic analysis to calculate the systematic errors
associated with the choice of basis, using
$\sigma_{b}=\sqrt{\frac{1}{N_{b}-1}\sum_{i=1}^{N_{b}}(M_{i}-\bar{M})^{2}}$,
where $N_{b}$=4 in this case.

In
Fig.~\ref{fig:m.avg_x1x1.6x6.sqrt.avgstaterr.basis_4combs.3rdTo6thCombs.5states.allQ},
the final results for the $6\times 6$ correlation matrix analysis
of $\chi_{1}\bar\chi_{1}$ correlators are shown. As the highest
excited state accommodates remaining spectral strength, this state
may not be an
eigenstate and is not shown in the figure. Masses
are averaged over the four bases (from the $3^{\rm rd}$ to
 $6^{\rm th}$), whereas the errors shown are a combination
of average statistical errors ($\bar{\sigma}_{s}$) over the four bases and systematic
errors due to the basis
choice ($\sigma_{b}$). As expected from the $4\times
4$ analysis~\cite{Mahbub:2009aa}, a similar lower lying Roper state
is also revealed in this $6\times 6$ correlation matrix analysis, which also has
a tendency to approach the physical state. This
analysis presents six distinct energy states, where the ground and 
first excited states are extremely robust, emphasizing the utility of the
analysis technique presented in Ref.~\cite{Mahbub:2009aa}. From this
figure it  is also  evident that
the second excited state remains high and an unnatural ordering of the
error bars is manifest. Hence there are concerns as to whether this
state is a true energy state.

 \begin{table*}
    \begin{center}
    \caption{\label{table:mpi.mavg.err-sum-over-qudrature.4comb.6states.x1x1}
      Masses of the nucleon, $N^{{\frac{1}{2}}^{+}}$ states, averaged over
      the four bases (from the $3^{\rm rd}$ to $6^{\rm th}$). The
      errors shown here are a combination of average statistical
      errors over four bases and systematic errors for the choice of
      basis, combined in quadrature.}
   \vspace{0.25cm}
    \begin{ruledtabular}
    \begin{tabular}{p{2cm}p{2cm}p{2cm}p{2cm}p{2cm}p{1.5cm}} 
   $aM_{\pi}$ & $aM_{g.s}$ & $aM_{1^{\rm st}\  \rm{e.s}}$(Roper) & $aM_{2^{\rm nd}\ \rm{e.s}}$ & $aM_{3^{\rm rd}\ \rm{e.s}}$ &$aM_{4^{\rm th}\ \rm{e.s}}$ \\
        \hline 
 0.5141(19) & 1.0414(68) & 1.451(40)   & 1.784(93)   & 2.447(60)   &  2.813(76)   \\   
 0.4705(20) & 0.9933(78) & 1.409(41)   & 1.892(98)   & 2.459(49)   &  2.856(78)   \\
 0.4134(22) & 0.9286(80) & 1.357(43)   & 1.875(54)   & 2.451(54)   &  2.838(90)   \\
 0.3490(24) & 0.8588(88) & 1.305(45)   & 1.857(61)   & 2.436(41)   &  2.849(59)   \\
 0.2776(24) & 0.781(10)  & 1.250(53)   & 1.886(65)   & 2.427(39)   &  2.845(54)   \\
 0.2452(24) & 0.752(11)  & 1.229(60)   & 1.903(38)   & 2.423(42)   &  2.846(58)   \\
 0.2110(27) & 0.722(14)  & 1.204(75)   & 1.894(42)   & 2.417(45)   &  2.850(72)   \\
 0.1905(31) & 0.711(12)  & 1.180(89)   & 1.892(46)   & 2.414(47)   &  2.840(74)   \\
 0.1607(35) & 0.682(14)  & 1.142(11)   & 1.891(51)   & 2.410(55)   &  2.837(72)   \\

  \end{tabular}
  \end{ruledtabular}
 \end{center}
 \end{table*}

\begin{figure*}
  \begin{center}
   $\begin{array}{c@{\hspace{0.15cm}}c}  
 \includegraphics [height=0.35\textwidth,angle=90]{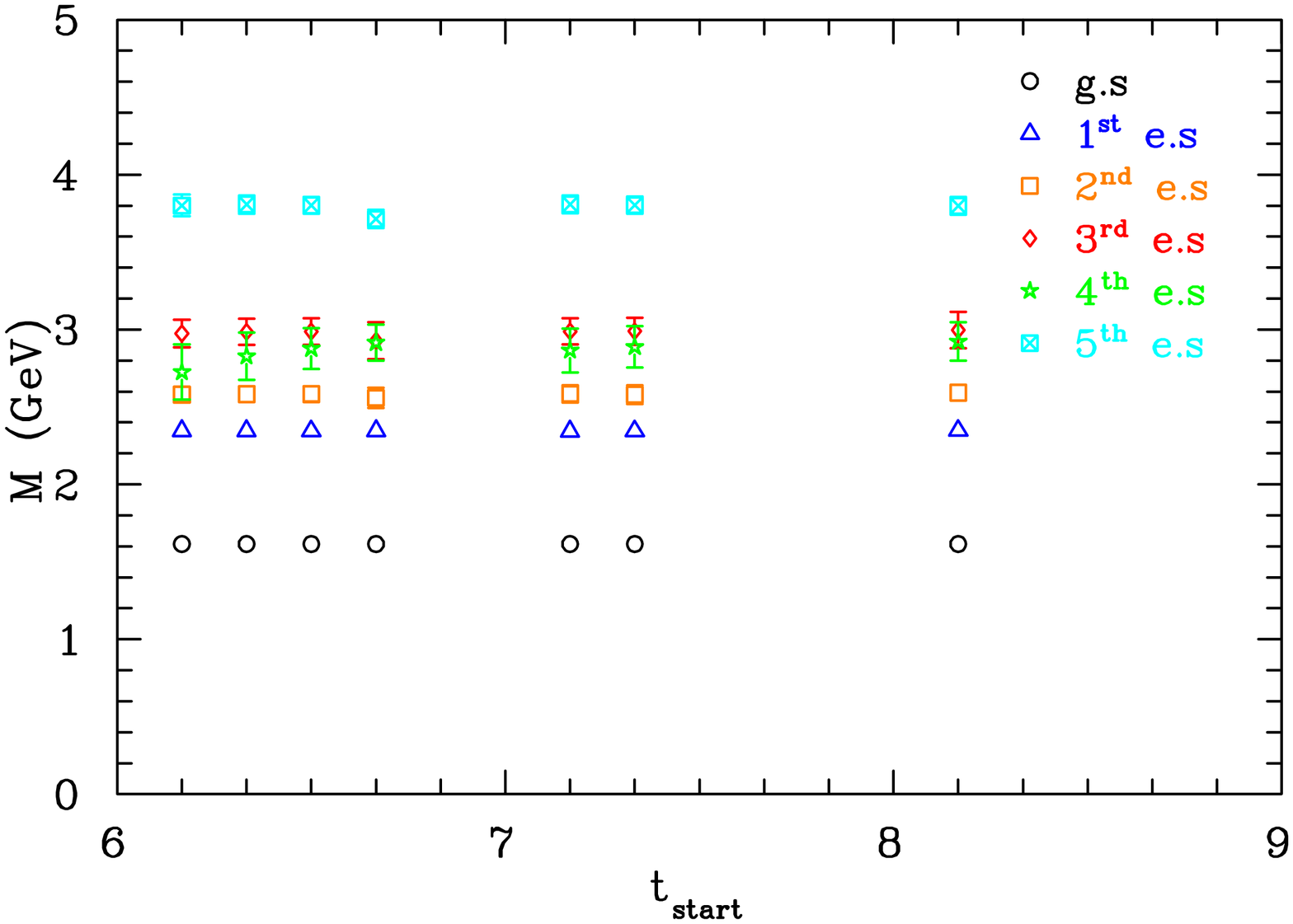} &
 \includegraphics [height=0.35\textwidth,angle=90]{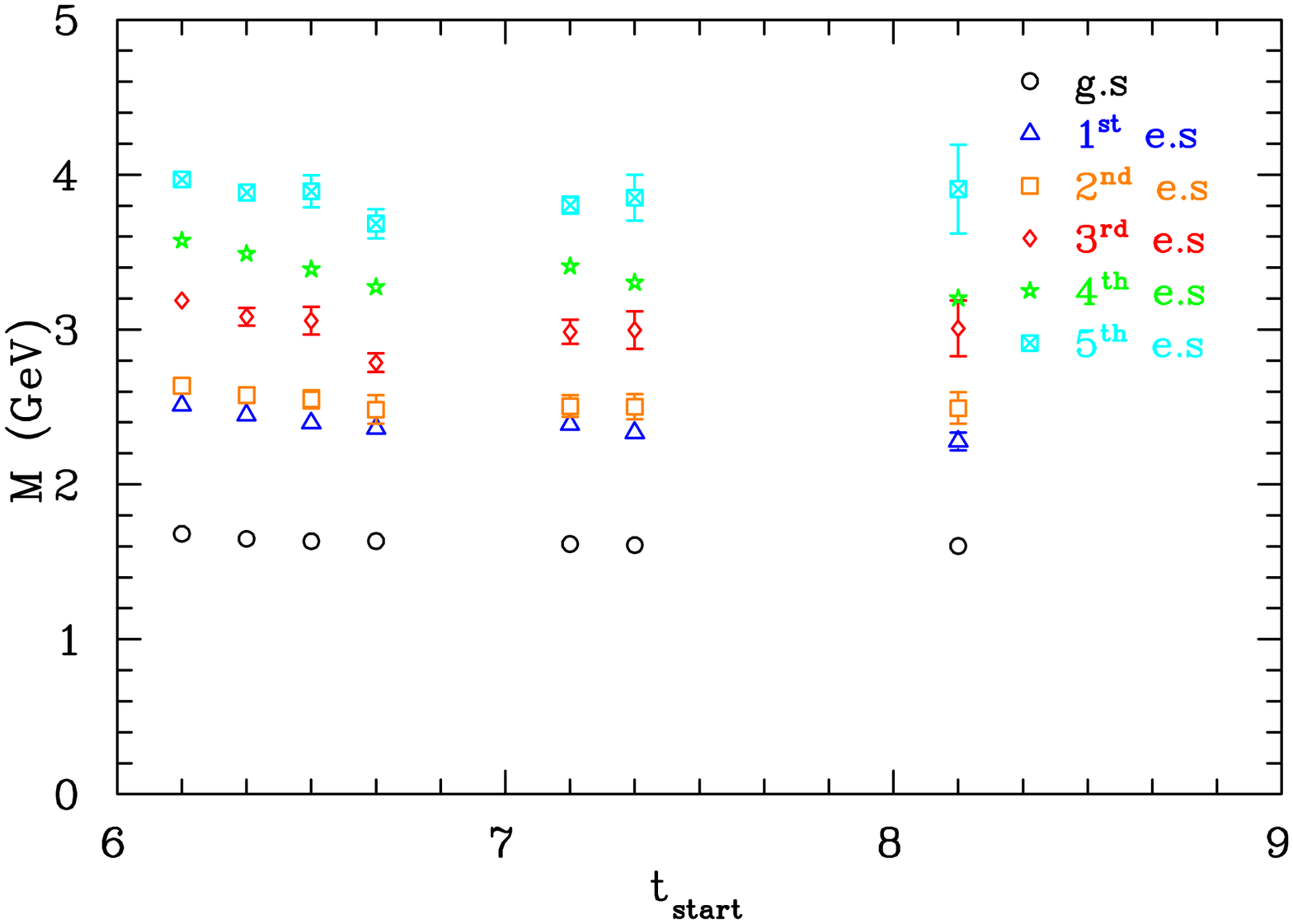}    
    \end{array}$

    \caption{(color online). Masses of the nucleon,
      $N^{{\frac{1}{2}}^{+}}$ states, from the projected correlation
      functions as shown in Eq.~\ref{eqn:projected_cf} (left) and  from
      the eigenvalues (right) for the $3^{\rm rd}$
      combination (3,16,48 sweeps) of $6\times 6$ correlation matrices of
      $\chi_{1}\chi_{2}$ correlation functions. The figure corresponds to
      the pion mass of 797 MeV. Each pair of ground and excited states
      mass correspond  to the
      diagonalization of the correlation matrix for each set of
      variational parameters $t_{\rm start}$ (shown in major tick marks) and
      $\triangle t$ (shown in minor tick marks).}   
   \label{fig:mass_and_eig_for_x1x2_6x6}  
  \end{center}
\end{figure*}

   In Ref.~\cite{Mahbub:2010me} addressing
   negative parity nucleons, a whole new set of nearby states is observed 
    upon introducing $\chi_{2}$. Thus, it is
   important to explore the role of $\chi_{2}$ in the positive-parity
   sector. In Fig.~\ref{fig:mass_and_eig_for_x1x2_6x6}, masses from the
projected correlation functions and eigenvalues for the 
$6\times 6$ correlation matrix analysis of $\chi_{1}\chi_{2}$
correlators are presented. The diagonalization is only successful for the heavier four
quarks, which again proves that a variational analysis with $\chi_{2}$
interpolator is always a challenge~\cite{Mahbub:2009nr}. This
interpolator vanishes in the non-relativistic
  limit~\cite{Leinweber:1990dv,Leinweber:1994nm} and is renowned for
  providing  noisier correlation
  functions~\cite{Sasaki:2001nf,Melnitchouk:2002eg} 
especially for the lighter quark masses, where one otherwise might
expect the relativistic nature of the quarks to be of benefit. The
Euclidean-time correlation functions of the $\chi_{2}$
interpolator decay more rapidly, and the signal-to-noise ratio
also deteriorates more quickly with time. In
Refs.~\cite{Mahbub:2009nr,Mahbub:2009aa}, a rise from below of excited 
 state masses from the projected correlation
functions was observed for the linearly dependent operators. This is
also manifest in the $6\times 6$ analysis of $\chi_{1}\bar\chi_{1}$, shown in
Fig.~\ref{fig:mass_and_eig_for_x1x1_6x6}, but for
the $6\times 6$ analysis of $\chi_{1}\chi_{2}$ correlators, these effects are
reduced. This is expected because the basis elements of these bases are
 likely to be more orthogonal, improving the isolation of states.


\begin{figure*}
  \begin{center}
   $\begin{array}{c@{\hspace{0.15cm}}c}  
 \includegraphics [height=0.40\textwidth,angle=90]{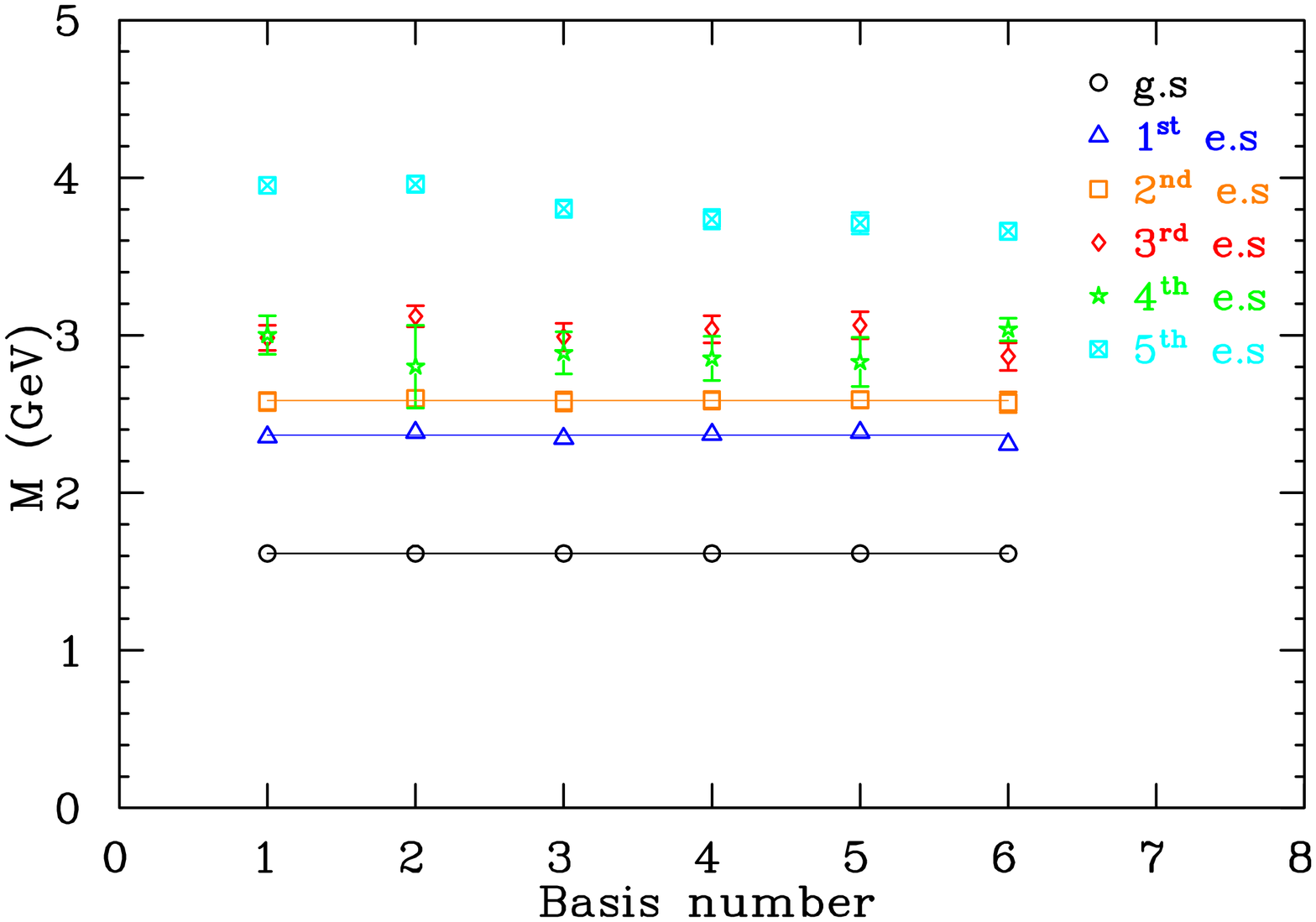} &
 \includegraphics [height=0.40\textwidth,angle=90]{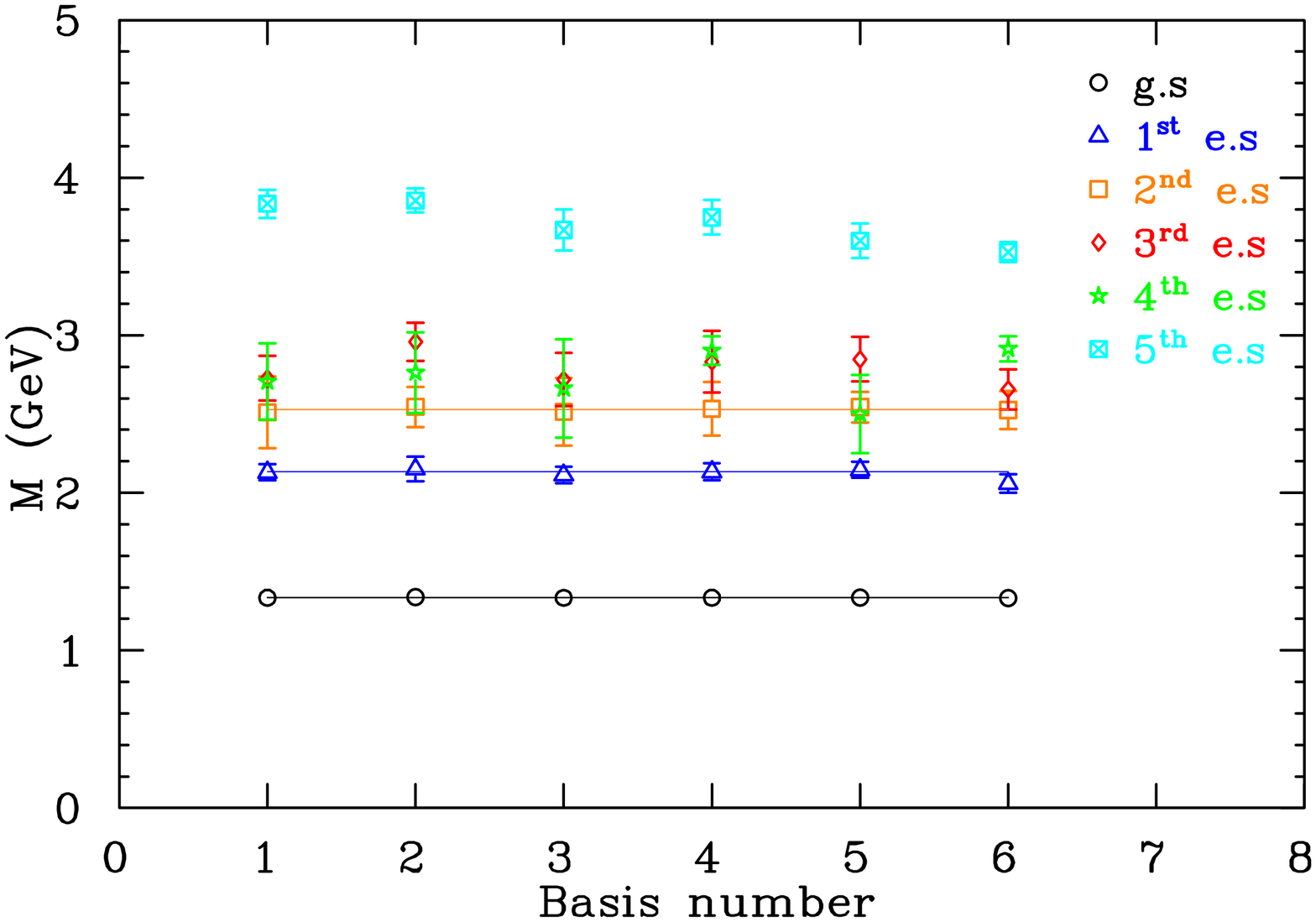}    
    \end{array}$
    \caption{(color online). Masses of the nucleon, $N^{{\frac
          {1}{2}}^{+}}$ states, from the projected correlation functions for
       the $6\times 6$ correlation matrices of $\chi_{1}\chi_{2}$
       correlators, for pion masses of 797 MeV (left) and 541 MeV
       (right).  Numbers on
      the horizontal scale correspond to each basis of $6\times 6$
      matrices, for instance, 1 and 2 correspond to the bases
      of (1,16,48 sweeps) and (3,12,26 sweeps) respectively, and so
      on. Masses are extracted according to the selection criteria described in
      the text and in Ref.~\cite{Mahbub:2009nr}
      from all the combinations of $6\times 6$ matrices as shown in
      Table~\ref{table:6x6_bases_x1x2}. Straight lines are drawn to
      illustrate the  invariance of
the masses over the bases.}   
   \label{fig:mass_6x6_x1x2_all_combinations}  
  \end{center}
\end{figure*}

In Fig.~\ref{fig:mass_6x6_x1x2_all_combinations}, projected masses from
all the $6\times 6$ correlation matrices of $\chi_{1}\chi_{2}$ are
presented. Again the ground state is robust in
this analysis. As the $6^{\rm th}$ combination contains consecutively
higher number of smearing sweeps (16,26,35), this basis is
not as reliable as the other sets. The first excited  state
extracted with this basis sits somewhat lower than the other five bases.
A careful
 analysis of Figs.~\ref{fig:mass_6x6_x1x2_all_combinations}
 and~\ref{fig:mass_6x6_x1x1_all_combinations} reveals that
  in the vicinity of the second excited state from
 the $6\times 6$ analysis of $\chi_{1}\bar\chi_{1}$ correlators 
(Fig.~\ref{fig:mass_6x6_x1x1_all_combinations}), three excited
states appear in the $6\times 6$ analysis of $\chi_{1}\chi_{2}$
(Fig.~\ref{fig:mass_6x6_x1x2_all_combinations}). A completely
 new excited state, the second excited state, is extracted in the $\chi_{1}\chi_{2}$
analysis. This second excited state is robust for all bases (left figure). The signal-to-noise ratio for this state deteriorates more rapidly
for the light quark (right figure). Interestingly, the fifth excited state
extracted from the $\chi_{1}\chi_{2}$ matrices, sits at a significantly lower
energy state  than those from the $\chi_{1}\bar\chi_{1}$ matrices,
reflecting the fact that two new low-lying states have been
revealed. The third and fourth
excited states are nearly degenerate in this analysis. As the
results for the lower three states are consistent over five basis
choices (from $1^{\rm st}$ to $5^{\rm th}$) of the $6\times 6$
matrices, we perform a systematic analysis over these five bases, as
discussed  previously.

\begin{figure*}
  \begin{center}

 \includegraphics [height=0.50\textwidth,angle=90]{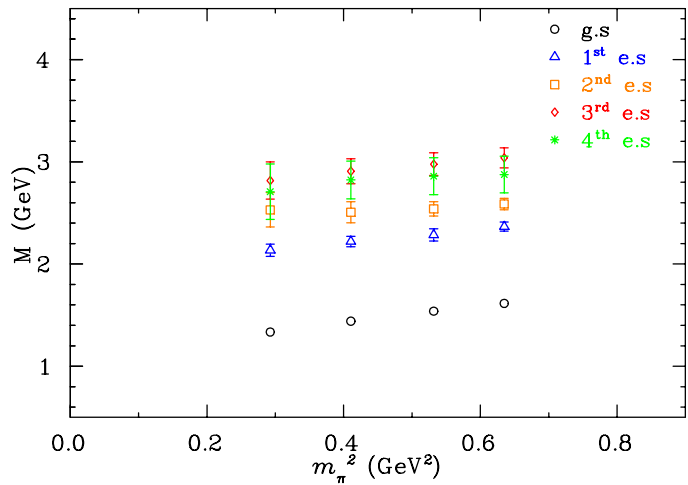}

    \caption{(color online). Masses of the nucleon,
      $N^{{\frac{1}{2}}^{+}}$ states, for the ground (g.s) and the
      excited (e.s) states from the
       $6\times 6$ correlation matrices of $\chi_{1}\chi_{2}$
      correlators over five bases (from the $1^{\rm st}$ to $5^{\rm th}$ 
      of Fig.~\ref{fig:mass_6x6_x1x2_all_combinations}). The
      errors shown in the figure are a combination of average statistical errors
       over these five bases and systematic errors due to basis
      choices. Errors are combined in quadrature. Masses are given in
      Table~\ref{table:mpi.mavg.err-sum-over-qudrature.5comb.5states.x1x2}.}  
   \label{fig:m.avg_x1x2.6x6.sqrt.avgstaterr.basis_5combs.5states.allQ}
  \end{center}
\end{figure*}

In Fig.~\ref{fig:m.avg_x1x2.6x6.sqrt.avgstaterr.basis_5combs.5states.allQ}, 
a summary of the $6\times 6$ correlation matrix analysis of
$\chi_{1}\chi_{2}$ correlators is presented. Masses are averaged over the
 five bases, while errors are a combination of average statistical
and systematic errors due to basis choices. Apart from the ground and
the first excited states, this analysis presents a distinct second
excited state for the three heavier quark masses. The signal for the
light quark mass for this state is poor. This
analysis  reveals new nearly degenerate third and fourth excited
states, that 
the $6\times 6$ bases of
$\chi_{1}\bar\chi_{1}$ correlation functions are unable to resolve.

 \begin{table*}
    \begin{center}
    \caption{\label{table:mpi.mavg.err-sum-over-qudrature.5comb.5states.x1x2}
      Masses of the nucleon, $N^{{\frac{1}{2}}^{+}}$ states, averaged over
      the five bases (from the $1^{\rm st}$ to $5^{\rm th}$). The
      errors shown here are a combination of average statistical
      errors over these bases and systematic errors for the choice of
      basis, combined in quadrature.}
   \vspace{0.5cm}
  \begin{ruledtabular}
    \begin{tabular}{p{2.5cm}p{2.5cm}p{2.5cm}p{2.5cm}p{2.5cm}p{1.25cm}} 

   $aM_{\pi}$ & $aM_{g.s}$ & $aM_{1^{\rm st}\  \rm{e.s}}$(Roper) & $aM_{2^{\rm nd}\ \rm{e.s}}$ & $aM_{3^{\rm rd}\ \rm{e.s}}$ &$aM_{4^{\rm th}\ \rm{e.s}}$ \\
        \hline 
 0.5141(19) & 1.0419(65)  & 1.526(28)  & 1.669(37) & 1.960(64) & 1.85(12) \\   
 0.4705(20) & 0.9927(69)  & 1.474(39)  & 1.638(45) & 1.920(72) & 1.84(12) \\
 0.4134(22) & 0.9295(75)  & 1.432(33)  & 1.617(66) & 1.875(79) & 1.82(12) \\
 0.3490(24) & 0.8612(84)  & 1.377(39)  & 1.63(11)  & 1.81(12)  & 1.74(17) \\

  \end{tabular}
  \end{ruledtabular}
 \end{center}
 \end{table*}


\begin{figure*}[!hpt]
  \begin{center}
   $\begin{array}{c@{\hspace{0.15cm}}c}  
 \includegraphics [height=0.35\textwidth,angle=90]{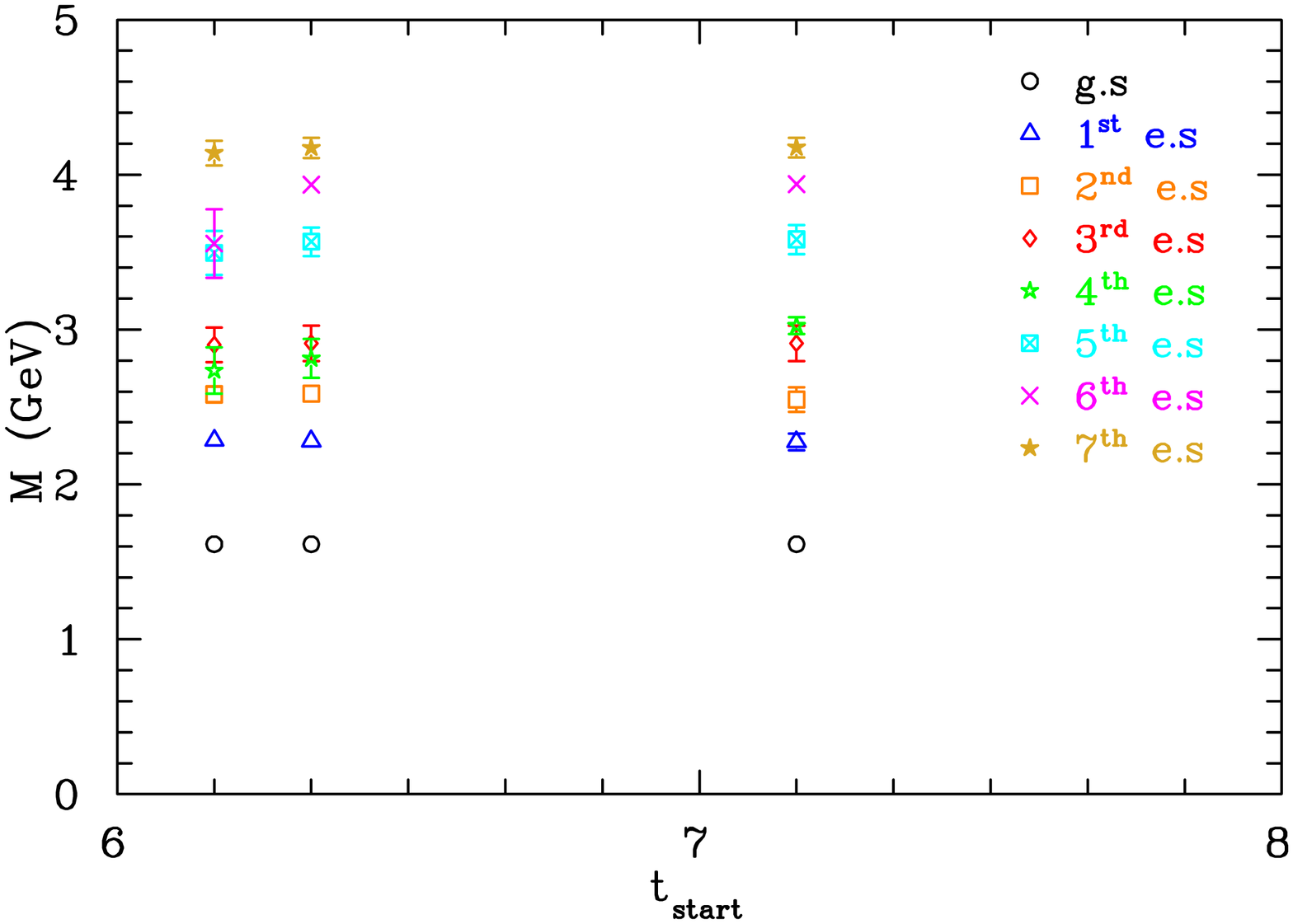} &
 \includegraphics [height=0.35\textwidth,angle=90]{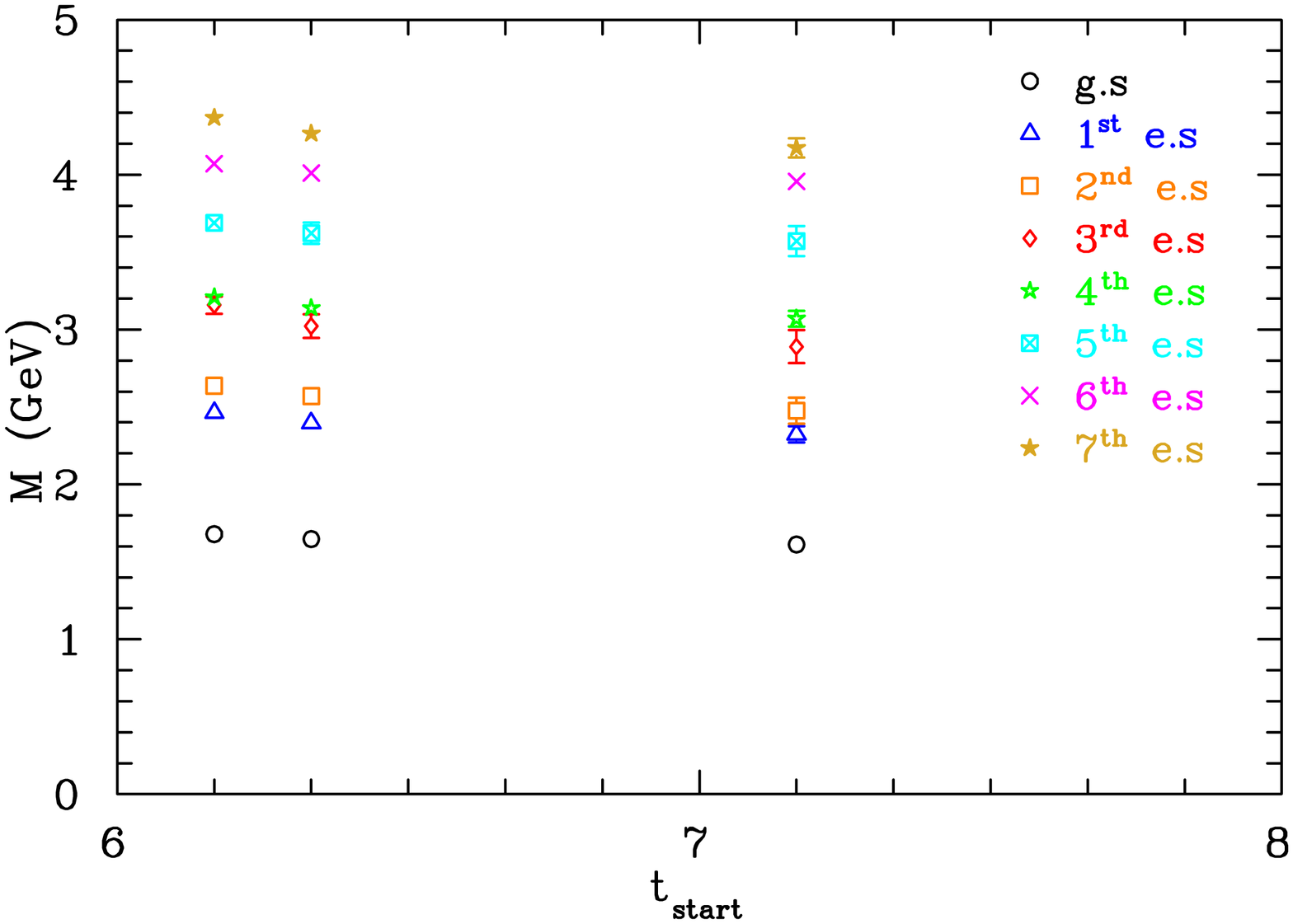}    
    \end{array}$

    \caption{(color online). Masses of the nucleon,
      $N^{{\frac{1}{2}}^{+}}$ states, from the projected correlation
      functions as shown in Eq.~\ref{eqn:projected_cf} (left) and  from
      the eigenvalues (right) for the $3^{\rm rd}$
      combination (3,16,48 sweeps) of $8\times 8$ correlation matrices of
      $\chi_{1}\chi_{2}$ correlation functions. The figure corresponds to
      the pion mass of 797 MeV. Each pair of ground and excited states
      mass correspond  to the
      diagonalization of the correlation matrix for each set of
      variational parameters $t_{\rm start}$ (shown in major tick marks) and
      $\triangle t$ (shown in minor tick marks).}   
   \label{fig:mass_and_eig_for_x1x2_8x8}  
  \end{center}
\end{figure*}

\begin{figure*}[!hpt]
  \begin{center}
   $\begin{array}{c@{\hspace{0.15cm}}c}  
 \includegraphics [height=0.40\textwidth,angle=90]{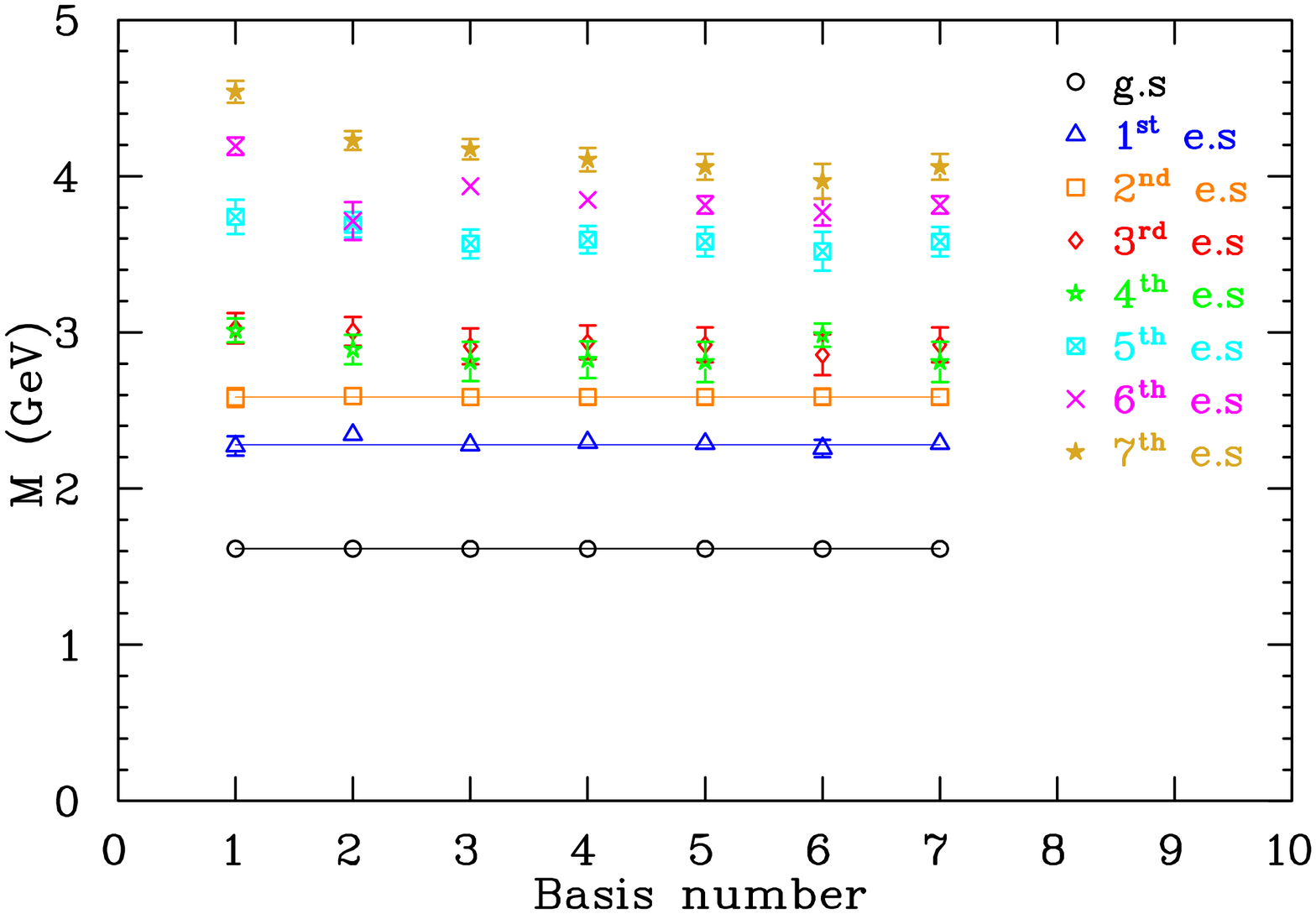} &
 \includegraphics [height=0.40\textwidth,angle=90]{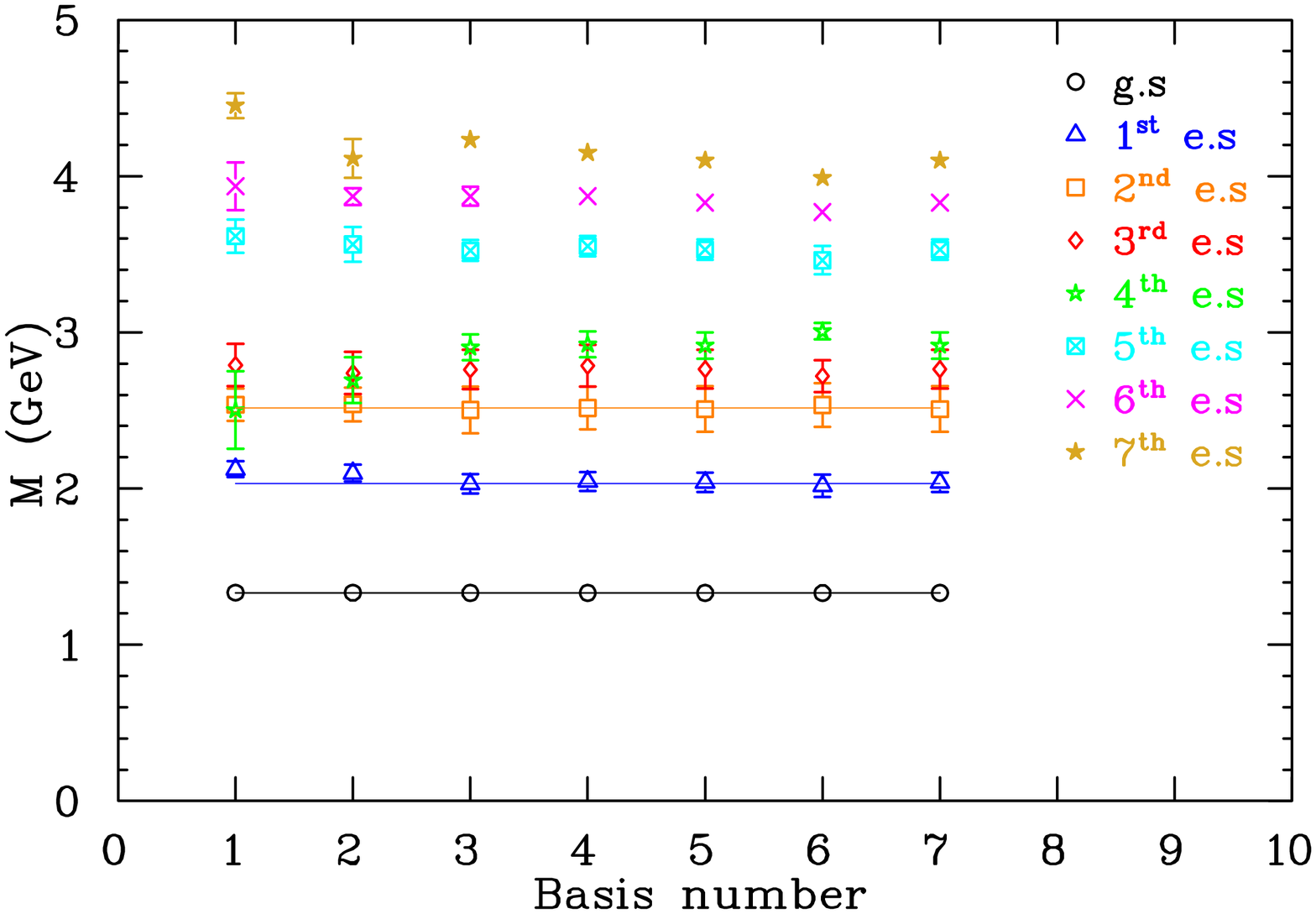}    
    \end{array}$
    \caption{(color online). Masses of the nucleon, $N^{{\frac
          {1}{2}}^{+}}$ states, from the projected correlation functions for
       the $8\times 8$ correlation matrices of $\chi_{1}\bar\chi_{2}$
       correlators, for pion masses of 797 MeV (left) and 541 MeV
       (right).  Numbers on
      the horizontal scale correspond to each basis of $8\times 8$
      matrices, for instance, 1 and 2 correspond to the bases
      of (1,7,16,35 sweeps) and (7,12,16,26 sweeps)
      respectively,  and so
      on. Masses are extracted according to the selection criteria described in
      the text and in Ref.~\cite{Mahbub:2009nr}
      from all the combinations of $8\times 8$ matrices as shown in
      Table~\ref{table:8x8_bases_x1x2}. Straight lines are drawn to
      illustrate the  invariance of
the masses over the bases.}   
   \label{fig:mass_8x8_x1x2_all_combinations}  
  \end{center}
\end{figure*}

\begin{figure*}[!hpt]
  \begin{center}

 \includegraphics [height=0.50\textwidth,angle=90]{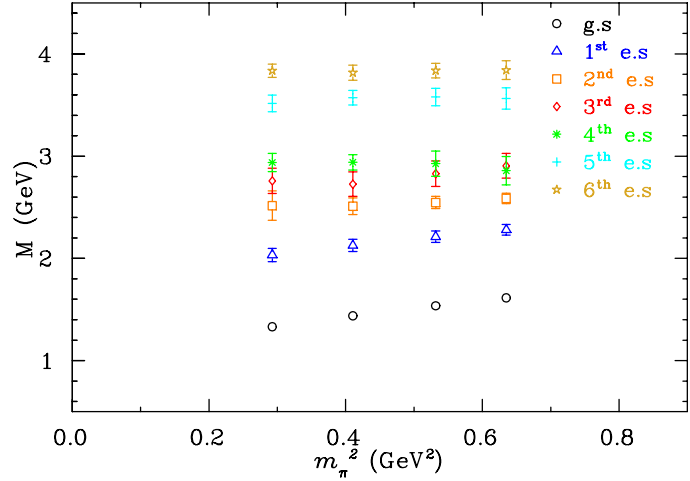} 
    \caption{(color online). Masses of the nucleon,
      $N^{{\frac{1}{2}}^{+}}$ states, for the ground (g.s) and the
      excited (e.s) states from the
       $8\times 8$ correlation matrices of $\chi_{1}\chi_{2}$
      correlators over four bases (from the $3^{\rm rd}$ to $6^{\rm th}$
      of Fig.~\ref{fig:mass_8x8_x1x2_all_combinations}). The
      errors shown in the figure are a combination of average statistical errors
       over these four bases and systematic errors due to basis
      choices, combined in quadrature. Masses are given in
      Table~\ref{table:mpi.mavg.err-sum-over-qudrature.4comb.7states.x1x2}.}  
   \label{fig:m.avg_x1x2.8x8.sqrt.avgstaterr.basis_4combs.7states.allQ}  
  \end{center}
\end{figure*}

 \begin{table*}
    \begin{center}
    \caption{\label{table:mpi.mavg.err-sum-over-qudrature.4comb.7states.x1x2}
      Masses of the nucleon, $N^{{\frac{1}{2}}^{+}}$ states, averaged over
      four bases (from the $3^{\rm rd}$ to $6^{\rm th}$). The
      errors shown here are a combination of average statistical
      errors over these four bases and systematic errors for the choice of
      basis, combined in quadrature.}
   \vspace{0.25cm}
    \begin{ruledtabular}
    \begin{tabular}{p{2.0cm}p{1.80cm}p{1.80cm}p{1.80cm}p{1.80cm}p{1.80cm}p{1.8cm} p{1.20cm}} 

   $aM_{\pi}$ & $aM_{g.s}$ & $aM_{1^{\rm st}\  \rm{e.s}}$(R) & $aM_{2^{\rm nd}\ \rm{e.s}}$ & $aM_{3^{\rm rd}\ \rm{e.s}}$ &$aM_{4^{\rm th}\ \rm{e.s}}$ & $aM_{5^{\rm th}\ \rm{e.s}}$ & $aM_{6^{\rm th}\ \rm{e.s}}$ \\
        \hline 
 0.5141(19) & 1.0412(66) & 1.470(33) & 1.669(33) & 1.874(78) & 1.843(90) & 2.300(67) & 2.478(58)  \\   
 0.4705(20) & 0.9912(71) & 1.427(35) & 1.642(38) & 1.825(79) & 1.888(80) & 2.309(54) & 2.475(46)  \\
 0.4134(22) & 0.9277(77) & 1.372(38) & 1.619(52) & 1.758(77) & 1.896(48) & 2.304(45) & 2.462(47)  \\
 0.3490(24) & 0.8588(86) & 1.311(42) & 1.623(92) & 1.779(80) & 1.895(57) & 2.269(52) & 2.475(41)  \\
  \end{tabular}
  \end{ruledtabular}
 \end{center}
 \end{table*}

In Fig.~\ref{fig:mass_and_eig_for_x1x2_8x8}, masses from
 the projected correlation functions and eigenvalues for the $8\times 8$
correlation matrix of $\chi_{1}\chi_{2}$ correlators are presented. Similar to the $6\times
6$ analysis of $\chi_{1}\chi_{2}$, this analysis is also successful for
the heavier four quark masses. The enhanced dimension of the matrices
means the numerical diagonalization is less stable, and so the
variational analysis is only successful for a few sets of variational parameters, $t_{\rm
  start}$ and $\triangle t$. However, in
Fig.~\ref{fig:mass_and_eig_for_x1x2_8x8}, there is sufficient consistency
 between the masses from the projected correlation functions for each
 set of $t_{\rm start}$ and $\triangle t$. As before, the $8\times 8$
 analysis gives a projected mass which is highly independent of the
 variational parameters. Here we have selected $t_{\rm start}=6$,
 $\triangle t = 2$.

In Fig.~\ref{fig:mass_8x8_x1x2_all_combinations}, masses from
 the projected correlation functions for all combinations of $8\times
8$  correlation matrices are shown. All the
bases reveal a very consistent ground state mass. A systematic basis
dependency of the excited states (from third excited state onwards) is
noticed for the $1^{\rm st}$ and $2^{\rm nd}$ combinations. As the
$1^{\rm st}$ combination consists of  1 and 7 smearing sweeps,
it provides a higher mass from the third excited state
onwards, in comparison with the other bases. The $2^{\rm nd}$ basis
contains consecutive smearing sweep counts, starting with 7 sweeps, 
and ending with a moderate sweep count of 26. This provides less
diversity in this basis for the higher excited states and also
provides a  higher mass from the third excited state
onwards. Therefore, the $1^{\rm st}$ and $2^{\rm nd}$ bases are not as reliable as other sets. The
fifth and  sixth excited states sit a little high for the $7^{\rm th}$
basis which has a narrow cluster of smearing sweeps and is therefore
likely to be less reliable in spanning the space.  
 Bases from the $3^{\rm rd}$ to $6^{\rm th}$ sets have more
 smearing diversity being well spread over the range of smearing sweep
 counts starting from 3 and
 7 sweeps. It is also evident from
 Fig.~\ref{fig:mass_8x8_x1x2_all_combinations} that the $3^{\rm rd}$
 to $6^{\rm th}$ combinations  of smearing sweeps are successful in
 providing highly consistent lower energy states. As before, we will
 therefore perform a systematic analysis
 over these four bases.  It is interesting to note that in the
 vicinity of  the fifth excited  state
 from the $6\times 6$ analysis of $\chi_{1}\chi_{2}$, the
 $8\times 8$ correlation matrix presents three 
 excited states.

In
Fig.~\ref{fig:m.avg_x1x2.8x8.sqrt.avgstaterr.basis_4combs.7states.allQ},
 a summary of the $8\times 8$ correlation matrix results are
presented. As in
Fig.~\ref{fig:m.avg_x1x2.6x6.sqrt.avgstaterr.basis_5combs.5states.allQ},
a distinct second excited state is also obtained in this $8\times 8$
analysis, for the two heavier quark masses. However, increased
separation of the third and fourth excited states appears here with
the enlarged basis. A  careful examination
reveals that the first excited state extracted in this analysis, is
 a little lower than in the $6\times 6$ analysis of
 $\chi_{1}\chi_{2}$ (see
 Fig.~\ref{fig:m.avg_x1x2.6x6.sqrt.avgstaterr.basis_5combs.5states.allQ}).
 In accord with the principle of the variational
method, our analysis signifies the importance of using larger
correlation matrices to reliably isolate the higher energy eigenstates of QCD.


\subsection{Splitting of Excited States}

\begin{figure}[!t]
 \begin{center} 
\includegraphics [height=0.48\textwidth,angle=90]{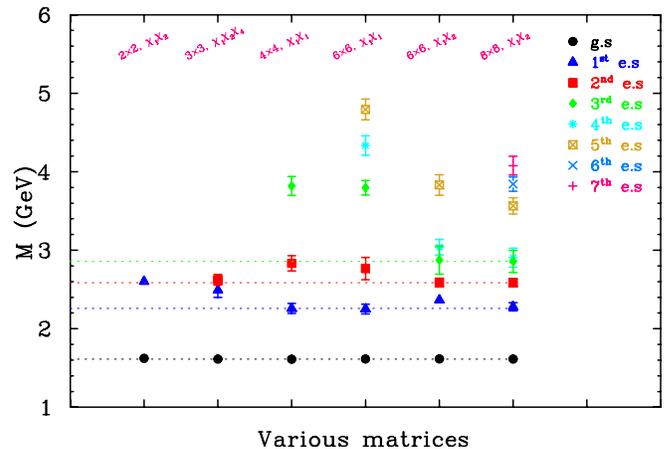}  
    \caption{(color online). Masses of the nucleon,
      $N^{{\frac{1}{2}}^{+}}$ states, for the various dimensions of
      correlation matrices, as labelled on the upper horizontal axis. The $2\times 2$
       and $3\times 3$ results for the smeared-smeared correlation
       functions with 26 sweeps of smearing are taken
       from  Ref.~\cite{Mahbub:2009nr}, while the
       $4\times 4$ results are from
       Ref.~\cite{Mahbub:2009aa}. The $6\times 6$
       and $8\times 8$ results correspond to 
       the analyses presented in this paper. Dotted lines are drawn to
       aid in  illustrating
       the consistency of the results. Figure corresponds to the
       pion mass of 797 MeV. } 
  \label{fig:2x2_3x3_4x4_6x6_6x6_8x8_history_FINAL}
 \end{center}
\end{figure}

In Fig.~\ref{fig:2x2_3x3_4x4_6x6_6x6_8x8_history_FINAL}, we show how the excited
energy states depend on the choice of correlation functions and
 the dimension of the correlation matrix. The ground
state is clearly independent of the dimension of the matrices and
choice of interpolating fields. The excited state extracted from the
$2\times 2$  correlation matrix splits into two excited
states with the $3\times 3$ matrix.
The central value of the first excited
state, extracted with the $3\times 3$ matrix, is a little lower than that of
$2\times 2$ case. Interestingly, the second excited state from the $3\times
3$ matrix is in very good agreement with the excited state number of the
$2\times 2$ analysis. 

The $4\times 4$ analysis of $\chi_{1}\bar\chi_{1}$ correlation
functions reveals a lower first excited state (Roper
state)~\cite{Mahbub:2009aa}  and two
other heavier excited states. Therefore, the $4\times 4$ basis of
$\chi_{1}\bar\chi_{1}$ correlation functions is able to resolve and isolate the 
superposition of eigenstates in the $3\times 3$ matrix. The $6\times 6$
and $4\times 4$ correlation matrices of $\chi_{1}\bar\chi_{1}$ correlators
provide very consistent results for the
lower four energy states, and the $6\times 6$ analysis is able to
extract two  new higher  energy
states. The larger error bar of the third (highest) excited state
extracted  with the $4\times 4$ correlation matrix compared to the
$6\times 6$ result of
$\chi_{1}\bar\chi_{1}$ is a manifestation of accommodating remaining
spectral strength in this (third excited) state in the $4\times 4$
analysis  and hence is
unreliable~\cite{Mahbub:2009aa}.

The $6\times 6$ analysis of
$\chi_{1}\chi_{2}$ correlation functions extracts different
excited states than those of the $6\times 6$ analysis of
$\chi_{1}\bar\chi_{1}$ alone. The first 
excited state extracted with this basis sits high in
comparison with the Roper state. Thus, a basis of four different
smearings with $\chi_{1}\bar\chi_{1}$ is essential to isolating the
single eigenstate associated with the Roper resonance. Another
interesting feature is the splitting of the ``second excited state''
of the $\chi_{1}\bar\chi_{1}$ analysis into three nearby states with
the new $\chi_{2}$ spin-flavour combinations.

 The final very interesting outcomes are revealed
from the $8\times 8$ correlation matrix analysis of $\chi_{1}\chi_{2}$. This
analysis provides a first excited state which is in excellent agreement with
the Roper state revealed by $\chi_{1}\bar\chi_{1}$ correlators
alone. This present investigation
   reveals that $\chi_{2}$ plays a subtle role for the
   Roper. We note however that a new nearby second excited state is
revealed in the $\chi_{1}\chi_{2}$ analysis and a future
high-statistics analysis may reveal it is essential to resolve this
state correctly to obtain the first excited state accurately. This basis also
provides second, third and fourth excited states consistent with the
$6\times 6$ analysis of $\chi_{1}\chi_{2}$.
 However, the fifth excited state coming from this analysis is a
 little lower  than that of
 the $6\times 6$ matrix of $\chi_{1}\chi_{2}$. There the fifth state
 accommodates all remaining spectral strength. This state wasn't
 identified in the $\chi_{1}\bar\chi_{1}$ analysis. The sixth excited
 state provided by this analysis is consistent with the third excited
 state extracted with the $6\times 6$ analysis of
 $\chi_{1}\bar\chi_{1}$. It  is worth noting that the basis of
$\chi_{1}\bar\chi_{1}$ correlation functions is insufficient to
 isolate the second excited state of the
 nucleon.

\begin{figure}[!t]
 \begin{center}
\includegraphics [height=0.48\textwidth,angle=90]{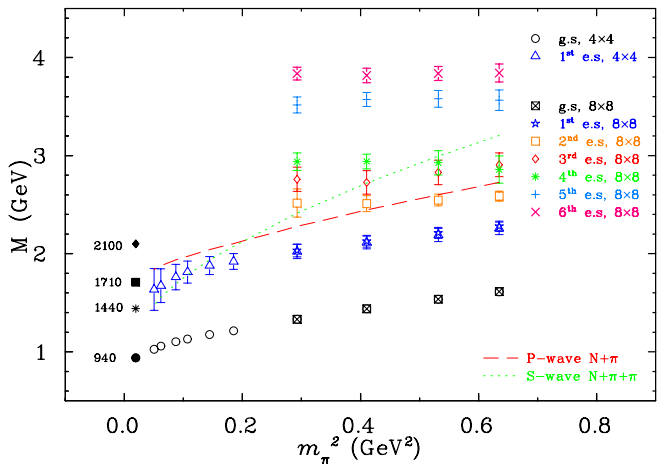}  
    \caption{(color online). Masses of the nucleon,
      $N^{{\frac{1}{2}}^{+}}$ states, from the $8\times 8$ correlation
      matrices of $\chi_{1}\chi_{2}$, and $4\times 4$ 
       correlation
      matrices of $\chi_{1}\bar\chi_{1}$ as a function of the squared pion
      mass, for the ground (g.s), first through sixth
      excited states (e.s).  The
      experimental values  are taken from
      Ref.~\cite{Amsler:2008zzb}, where the reliability of the ground and
      Roper  states are
      signified by four stars (****), ${\rm{P}}_{11}$ (1710 MeV) state
       by three stars (***) and ${\rm{P}}_{11}$ (2100 MeV) state
      by one star (*).}
      \label{fig:m.avg_x1x2.8x8.x1x1.4x4.7states.P_S_wave}
 \end{center}
\end{figure}

\begin{figure*}[!ht]
 \begin{center}
\includegraphics [height=0.95\textwidth,angle=90]{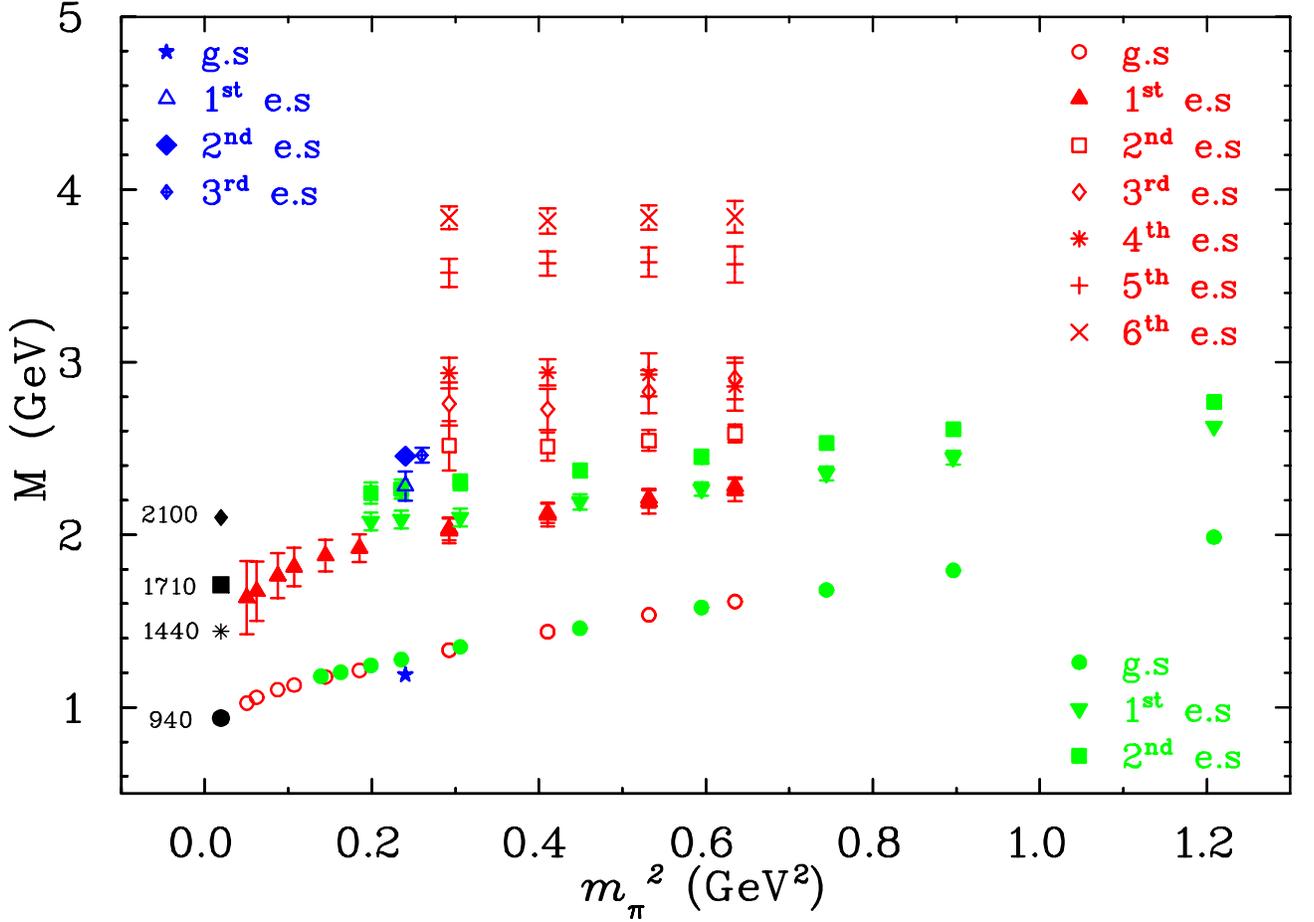}  
    \caption{(color online). Masses of low-lying nucleon states from
      this work (legend at top right), from
      Ref.~\cite{Basak:2007kj} (legend at
      top left) and  Ref.~\cite{Burch:2006cc} (legend at bottom
      right). The  results
        from Ref.~\cite{Basak:2007kj} correspond 
        to $m_{\pi}=490$ MeV, where
        the nearly degenerate fourth energy  state has been shifted
         a little to the right for clarity.} 
      \label{fig:m.avg_x1x2.8x8.x1x1.4x4.3groups}
 \end{center}
\end{figure*}

 While this  analysis is
able to provide some evidence that the first five energy-states are
reasonably robust through comparisons of $6\times 6$ and $8\times 8$,
only a more comprehensive higher dimension correlation matrix analysis
can assess the reliability of the sixth and seventh states.

In Fig.~\ref{fig:m.avg_x1x2.8x8.x1x1.4x4.7states.P_S_wave}, a
summary of results for the positive parity excited states of the
nucleon is  presented. In Ref.~\cite{Amsler:2008zzb}, the quality of
the ${\rm P}_{11}$ (1710) MeV state is
 characterized by three stars (***) and ${\rm P}_{11}$ (2100 MeV) state 
by one star (*). These resonance states decay  through the
$N\rightarrow N\pi\pi$ channel with positive parity. Therefore,
looking at both the S-wave $N+\pi+\pi$ and P-wave $N+\pi$ decay channels
provides information related to these resonances. The
second excited state is very close to the threshold $N+\pi$
state but has a different slope. It is also evident that this state is
significantly different
 and lower than the $N+\pi+\pi$ state for the heavier three quark
 masses providing evidence that this state is best described as a
   single particle state. It
   is also  noted that the third and fourth excited
   states have a mass dependence different from the
     multiparticle 
   states. This suggests that these two energy states revealed here are more
likely to be single particle  resonance states. Simulations on larger lattice
volumes and at lighter quark masses are necessary to resolve this issue.

Fig.~\ref{fig:m.avg_x1x2.8x8.x1x1.4x4.3groups} provides a comparison
of the  nucleon spectrum revealed in this
comprehensive analysis with state-of-the-art results from
Refs.~\cite{Basak:2007kj} and~\cite{Burch:2006cc} in quenched QCD. The
results  from the BGR collaboration~\cite{Burch:2006cc}
follow from an analysis similar to ours where fermion source smearing
is key to providing a basis which enables the isolation of excited
states.  However, they use only two levels of source smearing and use
different smearings for each quark flavour within a given
interpolator.  The fact that their Roper state sits high relative to
our result is consistent with our observation that a minimum of four
different smearing levels is required to obtain a stable low-lying
state.  Whereas our $6 \times 6$ analysis with six different smearing
levels did not change the mass of the Roper state from our $4 \times
4$ analysis, the consideration of only three smearing levels did lead
to a higher mass Roper state.

Results from the LHP collaboration~\cite{Basak:2007kj} are illustrated
at  $m_\pi = 490$ MeV.  Here we have reported all states that may have spin-1/2
quantum numbers. The Roper state in their
analysis sits  higher than
both the present analysis and the results of Ref.~\cite{Burch:2006cc}.
As their scale determination provides a low nucleon mass, the excited state
masses will become higher again if one sets the scale to reporoduce
the nucleon mass of this analysis or that of Ref.~\cite{Burch:2006cc}.
It suggests that their consideration of a single level of
source smearing may be insufficient to provide a basis suitable to resolve
the Roper state. Our expectation is that the introduction of a variety
  of source smearings into their basis
  operators may be beneficial. It would be interesting to expand their basis to
include several levels of source smearing to explore the extent to
which the excited state masses change.

\section{Conclusions}
\label{section:conclusions}

In this paper, we have presented a comprehensive correlation matrix
analysis of the positive parity spin-$\frac{1}{2}$ excitations of the
nucleon. We have considered large dimensions of correlation matrices,
built  from smeared-smeared correlation functions of 
$\chi_{1}\bar\chi_{1}$ and $\chi_{1}\chi_{2}$ correlators. The results of
this  paper, for the ground and Roper states of the nucleon, are very consistent
with the discovery of the Roper in Ref.~\cite{Mahbub:2009aa} from 
 a $4\times 4$ correlation matrix of $\chi_{1}\bar\chi_{1}$ alone.
Thus, the most important conclusion is that the
  earlier first  result for the Roper~\cite{Mahbub:2009aa} is robust
  under  the comprehensive analysis presented herein.

This paper signifies the importance of using a variety of
   smearings in making an operator basis and the importance of
   comparing the results across a range of choices of basis for the
     correlation matrix.  In particular, it emphasizes
   the use of at least four different smearing levels to isolate the elusive
   Roper state.

In addition, we have reported new second, third, fourth, fifth
 and sixth excited-states of the nucleon.
 While both the $\chi_{1}\bar\chi_{1}$ and $\chi_{1}\chi_{2}$ correlation
matrices provide excellent agreement for the ground and the Roper
states, only the $\chi_{1}\chi_{2}$ analysis provides access to a
completely new (second) excited eigenstate. All the
correlation matrix analyses provide a very consistent third excited
state, see Fig.~\ref{fig:2x2_3x3_4x4_6x6_6x6_8x8_history_FINAL}.
 Simulations on larger lattice volumes at lighter quark
masses  with higher
statistics will be interesting to investigate the propagation of
these states towards light quark masses.

Another interesting result of this paper is delineating the nature
of the splitting of the excited states of the nucleon, as seen in
 Fig.~\ref{fig:2x2_3x3_4x4_6x6_6x6_8x8_history_FINAL}, for the variational
method. While the ground state is independent of any basis choice, and
 the dimension of the matrix used, extracting a particular excited state
can depend on the interpolators making the correlation functions. For
 example, a variational analysis of $\chi_{1}\bar\chi_{1}$ correlation
functions doesn't provide the second excited state, indicating that
 the $\chi_{1}$ interpolator either doesn't couple or only has a very small
coupling to this state. We emphasize the importance of using a large
correlation matrix for a reliable extraction of excited states energies.

Because of the difficulties in ascertaining that a given set of basis
interpolators is sufficient to isolate all the eigenstates of QCD
in the mass range of interest, we
believe that it is essential to resolve the nucleon spectrum using a
wide variety of approaches. For example, it is now apparent that the
consideration of a single level of fermion source smearing is unlikely
to be sufficient to resolve the low-lying Roper state.  Future
investigations should explore the role of fermion source smearing in
providing a suite of basis operators that span the space in an effective
manner.

To test the robustness of the spectrum revealed in this analysis it
   will be important for future studies to carefully examine a wide
   variety of different interpolators in order to test whether or not
   other low-lying states will appear in the spectrum.  For example,
   we anticipate that five-quark operators will be essential in
   accessing the relevant low-lying meson-baryon scattering states.

\begin{acknowledgments}
This research was undertaken on the NCI National Facility in Canberra,
Australia, which is supported by the Australian Commonwealth
Government. We also acknowledge eResearch SA for generous
grants of supercomputing time which have enabled this project.  This
research is supported by the Australian Research Council.

\end{acknowledgments}


\begin{thebibliography}{38}
\expandafter\ifx\csname natexlab\endcsname\relax\def\natexlab#1{#1}\fi
\expandafter\ifx\csname bibnamefont\endcsname\relax
  \def\bibnamefont#1{#1}\fi
\expandafter\ifx\csname bibfnamefont\endcsname\relax
  \def\bibfnamefont#1{#1}\fi
\expandafter\ifx\csname citenamefont\endcsname\relax
  \def\citenamefont#1{#1}\fi
\expandafter\ifx\csname url\endcsname\relax
  \def\url#1{\texttt{#1}}\fi
\expandafter\ifx\csname urlprefix\endcsname\relax\def\urlprefix{URL }\fi
\providecommand{\bibinfo}[2]{#2}
\providecommand{\eprint}[2][]{\url{#2}}

\bibitem[{\citenamefont{Isgur and Karl}(1977)}]{Isgur:1977ef}
\bibinfo{author}{\bibfnamefont{N.}~\bibnamefont{Isgur}} \bibnamefont{and}
  \bibinfo{author}{\bibfnamefont{G.}~\bibnamefont{Karl}},
  \bibinfo{journal}{Phys. Lett.} \textbf{\bibinfo{volume}{B72}},
  \bibinfo{pages}{109} (\bibinfo{year}{1977}).

\bibitem[{\citenamefont{Isgur and Karl}(1979)}]{Isgur:1978wd}
\bibinfo{author}{\bibfnamefont{N.}~\bibnamefont{Isgur}} \bibnamefont{and}
  \bibinfo{author}{\bibfnamefont{G.}~\bibnamefont{Karl}},
  \bibinfo{journal}{Phys. Rev.} \textbf{\bibinfo{volume}{D19}},
  \bibinfo{pages}{2653} (\bibinfo{year}{1979}).

\bibitem[{\citenamefont{Li et~al.}(1992)\citenamefont{Li, Burkert, and
  Li}}]{Li:1991yba}
\bibinfo{author}{\bibfnamefont{Z.-p.} \bibnamefont{Li}},
  \bibinfo{author}{\bibfnamefont{V.}~\bibnamefont{Burkert}}, \bibnamefont{and}
  \bibinfo{author}{\bibfnamefont{Z.-j.} \bibnamefont{Li}},
  \bibinfo{journal}{Phys. Rev.} \textbf{\bibinfo{volume}{D46}},
  \bibinfo{pages}{70} (\bibinfo{year}{1992}).

\bibitem[{\citenamefont{Carlson and Mukhopadhyay}(1991)}]{Carlson:1991tg}
\bibinfo{author}{\bibfnamefont{C.~E.} \bibnamefont{Carlson}} \bibnamefont{and}
  \bibinfo{author}{\bibfnamefont{N.~C.} \bibnamefont{Mukhopadhyay}},
  \bibinfo{journal}{Phys. Rev. Lett.} \textbf{\bibinfo{volume}{67}},
  \bibinfo{pages}{3745} (\bibinfo{year}{1991}).

\bibitem[{\citenamefont{Guichon}(1985)}]{Guichon:1985ny}
\bibinfo{author}{\bibfnamefont{P.~A.~M.} \bibnamefont{Guichon}},
  \bibinfo{journal}{Phys. Lett.} \textbf{\bibinfo{volume}{B164}},
  \bibinfo{pages}{361} (\bibinfo{year}{1985}).

\bibitem[{\citenamefont{Krehl et~al.}(2000)\citenamefont{Krehl, Hanhart,
  Krewald, and Speth}}]{Krehl:1999km}
\bibinfo{author}{\bibfnamefont{O.}~\bibnamefont{Krehl}},
  \bibinfo{author}{\bibfnamefont{C.}~\bibnamefont{Hanhart}},
  \bibinfo{author}{\bibfnamefont{S.}~\bibnamefont{Krewald}}, \bibnamefont{and}
  \bibinfo{author}{\bibfnamefont{J.}~\bibnamefont{Speth}},
  \bibinfo{journal}{Phys. Rev.} \textbf{\bibinfo{volume}{{C62}}},
  \bibinfo{pages}{025207} (\bibinfo{year}{2000}), \eprint{nucl-th/9911080}.

\bibitem[{\citenamefont{Durr et~al.}(2008)}]{Durr:2008zz}
\bibinfo{author}{\bibfnamefont{S.}~\bibnamefont{Durr}} \bibnamefont{et~al.},
  \bibinfo{journal}{Science} \textbf{\bibinfo{volume}{322}},
  \bibinfo{pages}{1224} (\bibinfo{year}{2008}), \eprint{0906.3599}.

\bibitem[{\citenamefont{Leinweber}(1995)}]{Leinweber:1994nm}
\bibinfo{author}{\bibfnamefont{D.~B.} \bibnamefont{Leinweber}},
  \bibinfo{journal}{Phys. Rev.} \textbf{\bibinfo{volume}{D51}},
  \bibinfo{pages}{6383} (\bibinfo{year}{1995}), \eprint{nucl-th/9406001}.

\bibitem[{\citenamefont{Lee and Leinweber}(1999)}]{Lee:1998cx}
\bibinfo{author}{\bibfnamefont{F.~X.} \bibnamefont{Lee}} \bibnamefont{and}
  \bibinfo{author}{\bibfnamefont{D.~B.} \bibnamefont{Leinweber}},
  \bibinfo{journal}{Nucl. Phys. Proc. Suppl.} \textbf{\bibinfo{volume}{73}},
  \bibinfo{pages}{258} (\bibinfo{year}{1999}), \eprint{hep-lat/9809095}.

\bibitem[{\citenamefont{Gockeler et~al.}(2002)}]{Gockeler:2001db}
\bibinfo{author}{\bibfnamefont{M.}~\bibnamefont{Gockeler}} \bibnamefont{et~al.}
  (\bibinfo{collaboration}{QCDSF}), \bibinfo{journal}{Phys. Lett.}
  \textbf{\bibinfo{volume}{B532}}, \bibinfo{pages}{63} (\bibinfo{year}{2002}),
  \eprint{hep-lat/0106022}.

\bibitem[{\citenamefont{Sasaki et~al.}(2002)\citenamefont{Sasaki, Blum, and
  Ohta}}]{Sasaki:2001nf}
\bibinfo{author}{\bibfnamefont{S.}~\bibnamefont{Sasaki}},
  \bibinfo{author}{\bibfnamefont{T.}~\bibnamefont{Blum}}, \bibnamefont{and}
  \bibinfo{author}{\bibfnamefont{S.}~\bibnamefont{Ohta}},
  \bibinfo{journal}{Phys. Rev.} \textbf{\bibinfo{volume}{D65}},
  \bibinfo{pages}{074503} (\bibinfo{year}{2002}), \eprint{hep-lat/0102010}.

\bibitem[{\citenamefont{Melnitchouk et~al.}(2003)}]{Melnitchouk:2002eg}
\bibinfo{author}{\bibfnamefont{W.}~\bibnamefont{Melnitchouk}}
  \bibnamefont{et~al.}, \bibinfo{journal}{Phys. Rev.}
  \textbf{\bibinfo{volume}{D67}}, \bibinfo{pages}{114506}
  (\bibinfo{year}{2003}), \eprint{hep-lat/0202022}.

\bibitem[{\citenamefont{Edwards et~al.}(2003)\citenamefont{Edwards, Heller, and
  Richards}}]{Edwards:2003cd}
\bibinfo{author}{\bibfnamefont{R.~G.} \bibnamefont{Edwards}},
  \bibinfo{author}{\bibfnamefont{U.~M.} \bibnamefont{Heller}},
  \bibnamefont{and} \bibinfo{author}{\bibfnamefont{D.~G.}
  \bibnamefont{Richards}} (\bibinfo{collaboration}{LHP}),
  \bibinfo{journal}{Nucl. Phys. Proc. Suppl.} \textbf{\bibinfo{volume}{119}},
  \bibinfo{pages}{305} (\bibinfo{year}{2003}), \eprint{hep-lat/0303004}.

\bibitem[{\citenamefont{Lee et~al.}(2003)}]{Lee:2002gn}
\bibinfo{author}{\bibfnamefont{F.~X.} \bibnamefont{Lee}} \bibnamefont{et~al.},
  \bibinfo{journal}{Nucl. Phys. Proc. Suppl.} \textbf{\bibinfo{volume}{119}},
  \bibinfo{pages}{296} (\bibinfo{year}{2003}), \eprint{hep-lat/0208070}.

\bibitem[{\citenamefont{Mathur et~al.}(2005)}]{Mathur:2003zf}
\bibinfo{author}{\bibfnamefont{N.}~\bibnamefont{Mathur}} \bibnamefont{et~al.},
  \bibinfo{journal}{Phys. Lett.} \textbf{\bibinfo{volume}{B605}},
  \bibinfo{pages}{137} (\bibinfo{year}{2005}), \eprint{hep-ph/0306199}.

\bibitem[{\citenamefont{Sasaki}(2003)}]{Sasaki:2003xc}
\bibinfo{author}{\bibfnamefont{S.}~\bibnamefont{Sasaki}},
  \bibinfo{journal}{Prog. Theor. Phys. Suppl.} \textbf{\bibinfo{volume}{151}},
  \bibinfo{pages}{143} (\bibinfo{year}{2003}), \eprint{nucl-th/0305014}.

\bibitem[{\citenamefont{Basak et~al.}(2007)}]{Basak:2007kj}
\bibinfo{author}{\bibfnamefont{S.}~\bibnamefont{Basak}} \bibnamefont{et~al.},
  \bibinfo{journal}{Phys. Rev.} \textbf{\bibinfo{volume}{D76}},
  \bibinfo{pages}{074504} (\bibinfo{year}{2007}), \eprint{arXiv:0709.0008
  [hep-lat]}.

\bibitem[{\citenamefont{Bulava et~al.}(2009)}]{Bulava:2009jb}
\bibinfo{author}{\bibfnamefont{J.~M.} \bibnamefont{Bulava}}
  \bibnamefont{et~al.}, \bibinfo{journal}{Phys. Rev.}
  \textbf{\bibinfo{volume}{D79}}, \bibinfo{pages}{034505}
  (\bibinfo{year}{2009}), \eprint{0901.0027}.

\bibitem[{\citenamefont{Bulava et~al.}(2010)}]{Bulava:2010yg}
\bibinfo{author}{\bibfnamefont{J.}~\bibnamefont{Bulava}} \bibnamefont{et~al.},
  \bibinfo{journal}{Phys. Rev.} \textbf{\bibinfo{volume}{D82}},
  \bibinfo{pages}{014507} (\bibinfo{year}{2010}), \eprint{1004.5072}.

\bibitem[{\citenamefont{Engel et~al.}(2010)\citenamefont{Engel, Lang, Limmer,
  Mohler, and Schafer}}]{Engel:2010my}
\bibinfo{author}{\bibfnamefont{G.~P.} \bibnamefont{Engel}},
  \bibinfo{author}{\bibfnamefont{C.~B.} \bibnamefont{Lang}},
  \bibinfo{author}{\bibfnamefont{M.}~\bibnamefont{Limmer}},
  \bibinfo{author}{\bibfnamefont{D.}~\bibnamefont{Mohler}}, \bibnamefont{and}
  \bibinfo{author}{\bibfnamefont{A.}~\bibnamefont{Schafer}}
  (\bibinfo{collaboration}{BGR [Bern-Graz-Regensburg]}) (\bibinfo{year}{2010}),
  \eprint{1005.1748}.

\bibitem[{\citenamefont{Sasaki et~al.}(2005)\citenamefont{Sasaki, Sasaki, and
  Hatsuda}}]{Sasaki:2005ap}
\bibinfo{author}{\bibfnamefont{K.}~\bibnamefont{Sasaki}},
  \bibinfo{author}{\bibfnamefont{S.}~\bibnamefont{Sasaki}}, \bibnamefont{and}
  \bibinfo{author}{\bibfnamefont{T.}~\bibnamefont{Hatsuda}},
  \bibinfo{journal}{Phys. Lett.} \textbf{\bibinfo{volume}{B623}},
  \bibinfo{pages}{208} (\bibinfo{year}{2005}), \eprint{hep-lat/0504020}.

\bibitem[{\citenamefont{Michael}(1985)}]{Michael:1985ne}
\bibinfo{author}{\bibfnamefont{C.}~\bibnamefont{Michael}},
  \bibinfo{journal}{Nucl. Phys.} \textbf{\bibinfo{volume}{B259}},
  \bibinfo{pages}{58} (\bibinfo{year}{1985}).

\bibitem[{\citenamefont{Luscher and Wolff}(1990)}]{Luscher:1990ck}
\bibinfo{author}{\bibfnamefont{M.}~\bibnamefont{Luscher}} \bibnamefont{and}
  \bibinfo{author}{\bibfnamefont{U.}~\bibnamefont{Wolff}},
  \bibinfo{journal}{Nucl. Phys.} \textbf{\bibinfo{volume}{B339}},
  \bibinfo{pages}{222} (\bibinfo{year}{1990}).

\bibitem[{\citenamefont{Mahbub et~al.}(2009{\natexlab{a}})}]{Mahbub:2009aa}
\bibinfo{author}{\bibfnamefont{M.~S.} \bibnamefont{Mahbub}}
  \bibnamefont{et~al.}, \bibinfo{journal}{Phys. Lett.}
  \textbf{\bibinfo{volume}{B679}}, \bibinfo{pages}{418}
  (\bibinfo{year}{2009}{\natexlab{a}}), \eprint{0906.5433}.

\bibitem[{\citenamefont{Mahbub et~al.}(2009{\natexlab{b}})}]{Mahbub:2009nr}
\bibinfo{author}{\bibfnamefont{M.~S.} \bibnamefont{Mahbub}}
  \bibnamefont{et~al.}, \bibinfo{journal}{Phys. Rev.}
  \textbf{\bibinfo{volume}{{D80}}}, \bibinfo{pages}{054507}
  (\bibinfo{year}{2009}{\natexlab{b}}), \eprint{0905.3616}.

\bibitem[{\citenamefont{Gusken}(1990)}]{Gusken:1989qx}
\bibinfo{author}{\bibfnamefont{S.}~\bibnamefont{Gusken}},
  \bibinfo{journal}{Nucl. Phys. Proc. Suppl.} \textbf{\bibinfo{volume}{17}},
  \bibinfo{pages}{361} (\bibinfo{year}{1990}).

\bibitem[{\citenamefont{Blossier et~al.}(2009)\citenamefont{Blossier,
  Della~Morte, von Hippel, Mendes, and Sommer}}]{Blossier:2009kd}
\bibinfo{author}{\bibfnamefont{B.}~\bibnamefont{Blossier}},
  \bibinfo{author}{\bibfnamefont{M.}~\bibnamefont{Della~Morte}},
  \bibinfo{author}{\bibfnamefont{G.}~\bibnamefont{von Hippel}},
  \bibinfo{author}{\bibfnamefont{T.}~\bibnamefont{Mendes}}, \bibnamefont{and}
  \bibinfo{author}{\bibfnamefont{R.}~\bibnamefont{Sommer}},
  \bibinfo{journal}{JHEP} \textbf{\bibinfo{volume}{04}}, \bibinfo{pages}{094}
  (\bibinfo{year}{2009}), \eprint{0902.1265}.

\bibitem[{\citenamefont{Takaishi}(1996)}]{Takaishi:1996xj}
\bibinfo{author}{\bibfnamefont{T.}~\bibnamefont{Takaishi}},
  \bibinfo{journal}{Phys. Rev.} \textbf{\bibinfo{volume}{D54}},
  \bibinfo{pages}{1050} (\bibinfo{year}{1996}).

\bibitem[{\citenamefont{de~Forcrand et~al.}(2000)}]{deForcrand:1999bi}
\bibinfo{author}{\bibfnamefont{P.}~\bibnamefont{de~Forcrand}}
  \bibnamefont{et~al.} (\bibinfo{collaboration}{QCD-TARO}),
  \bibinfo{journal}{Nucl. Phys.} \textbf{\bibinfo{volume}{B577}},
  \bibinfo{pages}{263} (\bibinfo{year}{2000}), \eprint{hep-lat/9911033}.

\bibitem[{\citenamefont{Zanotti et~al.}(2002)}]{Zanotti:2001yb}
\bibinfo{author}{\bibfnamefont{J.~M.} \bibnamefont{Zanotti}}
  \bibnamefont{et~al.} (\bibinfo{collaboration}{CSSM Lattice}),
  \bibinfo{journal}{Phys. Rev.} \textbf{\bibinfo{volume}{D65}},
  \bibinfo{pages}{074507} (\bibinfo{year}{2002}), \eprint{hep-lat/0110216}.

\bibitem[{\citenamefont{Zanotti et~al.}(2005)\citenamefont{Zanotti, Lasscock,
  Leinweber, and Williams}}]{Zanotti:2004dr}
\bibinfo{author}{\bibfnamefont{J.~M.} \bibnamefont{Zanotti}},
  \bibinfo{author}{\bibfnamefont{B.}~\bibnamefont{Lasscock}},
  \bibinfo{author}{\bibfnamefont{D.~B.} \bibnamefont{Leinweber}},
  \bibnamefont{and} \bibinfo{author}{\bibfnamefont{A.~G.}
  \bibnamefont{Williams}}, \bibinfo{journal}{Phys. Rev.}
  \textbf{\bibinfo{volume}{D71}}, \bibinfo{pages}{034510}
  (\bibinfo{year}{2005}), \eprint{hep-lat/0405015}.

\bibitem[{\citenamefont{Sommer}(1994)}]{Sommer:1993ce}
\bibinfo{author}{\bibfnamefont{R.}~\bibnamefont{Sommer}},
  \bibinfo{journal}{Nucl. Phys.} \textbf{\bibinfo{volume}{B411}},
  \bibinfo{pages}{839} (\bibinfo{year}{1994}), \eprint{hep-lat/9310022}.

\bibitem[{\citenamefont{Morningstar and Peardon}(2004)}]{Morningstar:2003gk}
\bibinfo{author}{\bibfnamefont{C.}~\bibnamefont{Morningstar}} \bibnamefont{and}
  \bibinfo{author}{\bibfnamefont{M.~J.} \bibnamefont{Peardon}},
  \bibinfo{journal}{Phys. Rev.} \textbf{\bibinfo{volume}{D69}},
  \bibinfo{pages}{054501} (\bibinfo{year}{2004}), \eprint{hep-lat/0311018}.

\bibitem[{\citenamefont{Lasscock et~al.}(2005)}]{Lasscock:2005kx}
\bibinfo{author}{\bibfnamefont{B.~G.} \bibnamefont{Lasscock}}
  \bibnamefont{et~al.}, \bibinfo{journal}{Phys. Rev.}
  \textbf{\bibinfo{volume}{D72}}, \bibinfo{pages}{074507}
  (\bibinfo{year}{2005}), \eprint{hep-lat/0504015}.

\bibitem[{\citenamefont{Amsler et~al.}(2008)}]{Amsler:2008zzb}
\bibinfo{author}{\bibfnamefont{C.}~\bibnamefont{Amsler}} \bibnamefont{et~al.}
  (\bibinfo{collaboration}{Particle Data Group}), \bibinfo{journal}{Phys.
  Lett.} \textbf{\bibinfo{volume}{B667}}, \bibinfo{pages}{1}
  (\bibinfo{year}{2008}).


\bibitem[{\citenamefont{Mahbub et~al.}(2010)\citenamefont{Mahbub, Kamleh,
  Leinweber, Cais, and Williams}}]{Mahbub:2010me}
\bibinfo{author}{\bibfnamefont{M.~S.} \bibnamefont{Mahbub}},
  \bibinfo{author}{\bibfnamefont{W.}~\bibnamefont{Kamleh}},
  \bibinfo{author}{\bibfnamefont{D.~B.} \bibnamefont{Leinweber}},
  \bibinfo{author}{\bibfnamefont{A.~O.} \bibnamefont{Cais}}, \bibnamefont{and}
  \bibinfo{author}{\bibfnamefont{A.~G.} \bibnamefont{Williams}},
  \bibinfo{journal}{Phys. Lett.} \textbf{\bibinfo{volume}{B693}},
  \bibinfo{pages}{351} (\bibinfo{year}{2010}), \eprint{1007.4871}.


\bibitem[{\citenamefont{Leinweber et~al.}(1991)\citenamefont{Leinweber,
  Woloshyn, and Draper}}]{Leinweber:1990dv}
\bibinfo{author}{\bibfnamefont{D.~B.} \bibnamefont{Leinweber}},
  \bibinfo{author}{\bibfnamefont{R.~M.} \bibnamefont{Woloshyn}},
  \bibnamefont{and} \bibinfo{author}{\bibfnamefont{T.}~\bibnamefont{Draper}},
  \bibinfo{journal}{Phys. Rev.} \textbf{\bibinfo{volume}{D43}},
  \bibinfo{pages}{1659} (\bibinfo{year}{1991}).

\bibitem[{\citenamefont{Burch et~al.}(2006)}]{Burch:2006cc}
\bibinfo{author}{\bibfnamefont{T.}~\bibnamefont{Burch}} \bibnamefont{et~al.},
  \bibinfo{journal}{Phys. Rev.} \textbf{\bibinfo{volume}{D74}},
  \bibinfo{pages}{014504} (\bibinfo{year}{2006}), \eprint{hep-lat/0604019}.

\end{thebibliography}
\end{document}